\documentclass[12pt]{article}
\usepackage{amsmath}
\usepackage{datetime}
\usepackage{amssymb} 
\usepackage{graphicx}
\usepackage{enumerate,color}
\usepackage{url}
\usepackage[pdfborder={0 0 0}]{hyperref}
\usepackage{appendix}
\usepackage{eso-pic}
\usepackage{amsmath}
\usepackage{amssymb}
\usepackage[nottoc]{tocbibind}
\usepackage[authoryear]{natbib}
\usepackage[sf,normalsize]{subfigure}
\usepackage[format=plain,labelformat=simple,labelsep=colon]{caption}
\usepackage{placeins}
\usepackage{tabularx}
\usepackage{morefloats}
\usepackage{stackengine}
\usepackage{framed}
\usepackage{minitoc}
\usepackage{color}
\usepackage{algorithm}
\usepackage{algorithmic}
\usepackage{bm}
\usepackage{multirow}
\usepackage{ifetex}
\usepackage{subfigure}
\usepackage[T1]{fontenc}
\usepackage[utf8]{inputenc}
\graphicspath{{figs/}}
\newcommand{\gumbel}{\text{Gumbel}}
\newcommand{\blind}{0}

\begin{document}

\def\spacingset#1{\renewcommand{\baselinestretch}%
{#1}\small\normalsize} \spacingset{1}

%%%%%%%%%%%%%%%%%%%%%%%%%%%%%%%%%%%%%%%%%%%%%%%%%%%%%%%%%%%%%%%%%%%%%%%%%%%%%%

\if0\blind
{
  \title{\bf A Bayesian hierarchical model for monthly maxima of
    instantaneous flow}
  \author{Egil Ferkingstad\\
    Science Institute, University of Iceland\\
    and\\
    Oli Pall Geirsson\\
     Science Institute, University of Iceland\\
    and\\
    Birgir Hrafnkelsson\\
    Faculty of Physical Sciences,
    University of Iceland\\
    and \\
    Olafur Birgir Davidsson\\
    deCODE genetics\\
    and \\   
    Sigurdur Magnus Gardarsson\\
    Faculty of Environmental and Civil Engineering, University of Iceland   
  }
  \maketitle
} \fi

\if1\blind
{
  \bigskip
  \bigskip
  \bigskip
  \begin{center}
    {\LARGE\bf A Bayesian hierarchical model for monthly maxima of
    instantaneous flow}
\end{center}
  \medskip
} \fi

\bigskip
\begin{abstract}
We propose a comprehensive Bayesian hierarchical model for monthly
maxima of instantaneous flow in river catchments.
The Gumbel distribution is used as the probabilistic model for the 
observations, which are assumed to come from several
catchments.
Our suggested latent model is Gaussian and designed
for monthly maxima, making better use of the
data than the standard approach using annual maxima.
At the latent level, linear mixed  models are used for both
the location and scale parameters of the Gumbel distribution,
accounting for seasonal dependence and covariates from the
catchments. 
The specification of prior distributions makes use
of penalised complexity (PC) priors, to ensure robust inference 
for the latent parameters. The main idea behind the
PC priors is to shrink toward a base model, thus avoiding
overfitting. PC priors also provide a convenient framework for
prior elicitation based on simple notions of scale. Prior
distributions for regression coefficients are also elicited based on
hydrological and meteorological knowledge.
Posterior inference was done using the MCMC split sampler, an efficient
Gibbs blocking scheme tailored to
latent Gaussian models. 
The proposed model was applied to observed data from eight river catchments in Iceland.
A cross-validation study demonstrates good predictive performance. 

\end{abstract}

\noindent%
\vfill

\newpage
\spacingset{1.45} % DON'T change the spacing!

\section{Introduction}

\label{sec:intro}

A common and substantial problem in hydrology is that of estimating the return period of extreme floods. An accurate estimate of extreme floods is of interest in various circumstances, particularly with respect to important civil infrastructure.  The design and construction of bridges and roads is often dependent on accurate understanding of river behavior during extreme events.  Changes in land use, especially in the urban environment, create increasingly more impervious surfaces. This leads to larger and more frequent floods, putting more stresses on flood control structures, such as levees and dams.  Climate change alters local precipitation patterns and magnitudes. This influences water resource management of reservoirs and rivers, affecting operation of hydroelectric power plants and river transport.  The management, operation, and maintenance of this critical infrastructure relies on accurate flood predictions, including predictions for ungauged catchments based on data from gauged river catchments.

One of the first approaches to regional flood estimation was the \emph{index flood method}, first proposed by \cite{dalrymple1960flood}. It was designed to deal with cases where little or no at-site data is available for flood assessment by borrowing strength from similar (e.g.~neighboring) gauged catchments. The method consists of two main steps, namely, regionalization, which includes the identification of geographically and climatologically homogeneous regions, and 
the specification of a regional standardized flood frequency curve for a $T$-year return period. In Section~\ref{sec:model} a mathematical formalization of the index flood method is used to motivate some of the elements of our proposed model. 

The index flood method is still widely used today, and further developments of the method were presented in~\cite{hosking1985estimation} and~\cite{grehy1996presentation}. Starting with the work of~\cite{cunnane1974bayesian}, various Bayesian extensions have been proposed~\citep{rosbjerg1995uncertainty, kuczera1999comprehensive, martins2000generalized}. Although these papers show the usefulness of Bayesian methods, they all derive rather directly from the classical index flood method, their main goal is usually to improve the estimation of the index flood coefficient, and they all rely solely on annual maxima. 
This work improves on the above studies in many important ways: The power relationship used to estimate the index flood coefficients is instead employed in the priors for the parameters of the Gumbel distribution, which we have chosen as the distribution for the observations. We use carefully chosen meteorological and topographical covariates, including  catchment areas and covariates based on precipitation and temperature measurements, motivated by the work of~\cite{crochet2012estimating}.
In summary, we believe that our work provides a coherent and comprehensive Bayesian model, making better use of the available data and prior knowledge.

We propose a Bayesian hierarchical model for monthly instantaneous extreme flow data from several river catchments. The topographical and climatic covariates facilitate the process of extrapolating the model to ungauged river catchments. 
Several novelties in statistical modeling and inference for flood data are presented here:
 We use monthly rather than yearly maxima, making better use of the available data. We use  a latent Gaussian model (LGM, see e.g.~\citet{rue2009approximate}) incorporating seasonal dependence, borrowing strength across months. The LGM allows the use of the computationally efficient MCMC split samling  algorithm~\citep{geirsson2015mcmc}, while still being sufficiently general to allow for realistic modeling.
We use  penalised-complexity priors~\citep{simpson2014penalising} for the hyperparameters of the model, which avoids overfitting, letting the prior knowledge together with the data decide the appropriate level of model complexity. We do a thorough prior eliciation for the regression coefficients of our model, making good use of availiable prior knowledge. To demonstrate that the proposed model predicts well for ungauged catchments, we perform a cross-validation study, where we leave river $j$ out and predict based on the model estimated from the other rivers except $j$, for each of the eight rivers.

We proceed as follows:
Section~\ref{sec:data} presents the data and the hydrological aspects of the problem. 
Section~\ref{sec:model} introduces the full hierarchical model and provides explanations of the modelling assumptions, and a description of the posterior inference.  Section~\ref{sec:results} summarizes the results obtained from applying the model to the data. Finally, Section~\ref{sec:conclusion} contains the conclusions drawn from the study and some ideas for future research.

%\newpage
\section{Data}
\label{sec:data}

\subsection{Streamflow Data and River Catchments}

\begin{figure}[!h]
\centering
\includegraphics[scale=0.6]{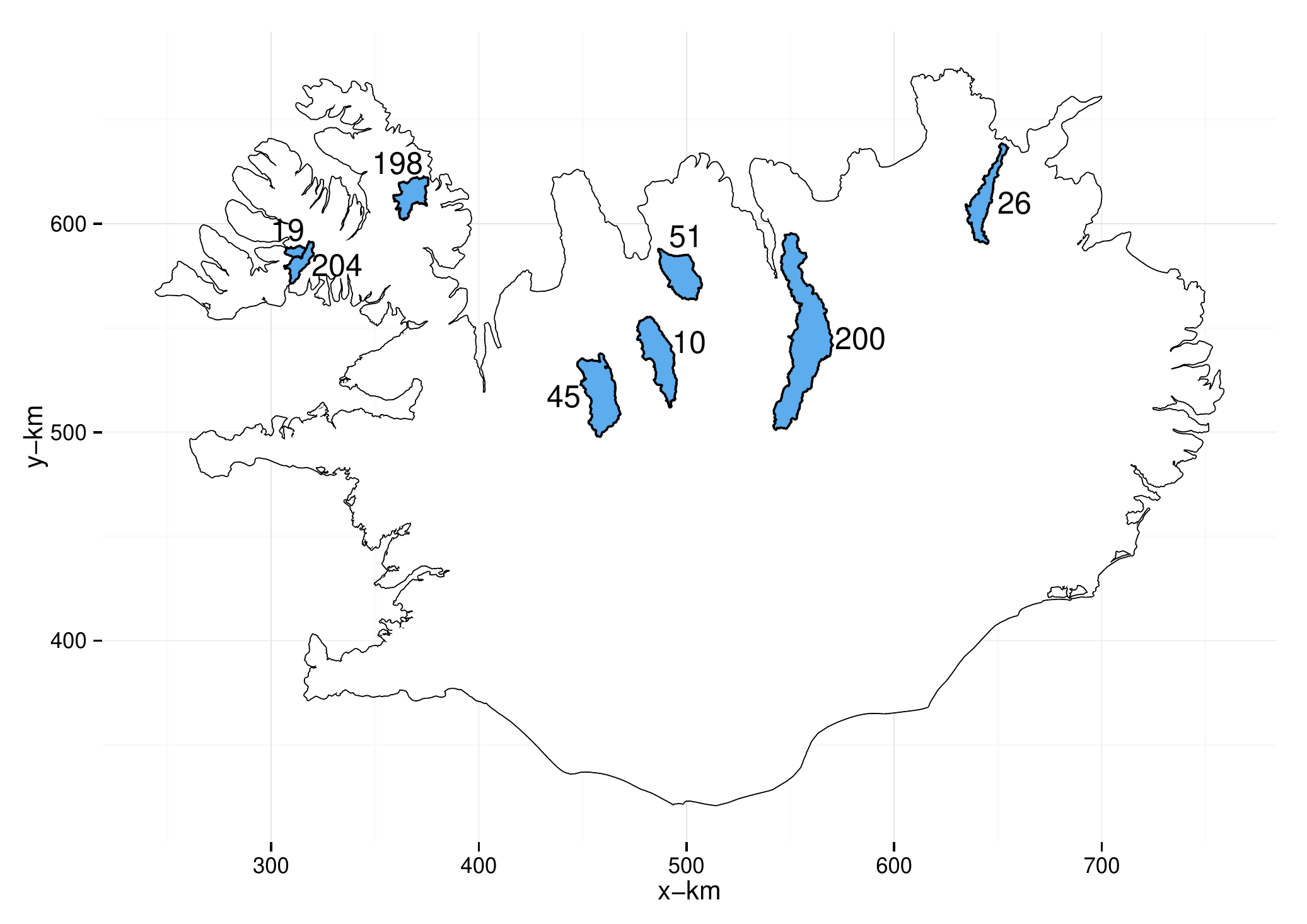}
\caption{Locations of river catchments. Catchment boundaries are provided by the Icelandic Meteorological Office. Coastline is provided by Landsvirkjun, the National Power Company of Iceland.} 
\label{fig:iceland}
\end{figure}

The streamflow data  consist of monthly maximum instantaneous discharges from eight river catchments in Iceland.  Table \ref{stationtable} lists the identification number, name and the size of each catchment. 
Even though stations VHM45 and VHM204 have the same name (Vatnsdalsa), they correspond to different catchments.
The time series were between 20 and 80 years long (in most cases between 40 and 60 years).  Figure~\ref{fig:iceland} shows the locations of the eight catchments.

\begin{table}[H]
\centering
\caption{Characteristics of the catchments used in the study.  The station identifications, river names and catchment areas were provided by the Icelandic Meteorological Office. }
\begin{tabular}{llc|llc}
\hline
Station & River & Area ($\text{km}^2$) & Station & River & Area ($\text{km}^2$)\\ 
\hline
VHM10 & Svarta & 392  & VHM51 & Hjaltadalsa & 296 \\
VHM19 & Dynjandisa & 37 & VHM198 & Hvala & 195 \\ 
VHM26 & Sanda & 267  &  VHM200 & Fnjoska & 1094 \\ 
VHM45 & Vatnsdalsa & 456  & VHM204 & Vatnsdalsa  & 103 \\ 
\hline
\end{tabular} 
\label{stationtable}
\end{table}

\begin{figure}[!h]
\centering
\includegraphics[scale=0.75]{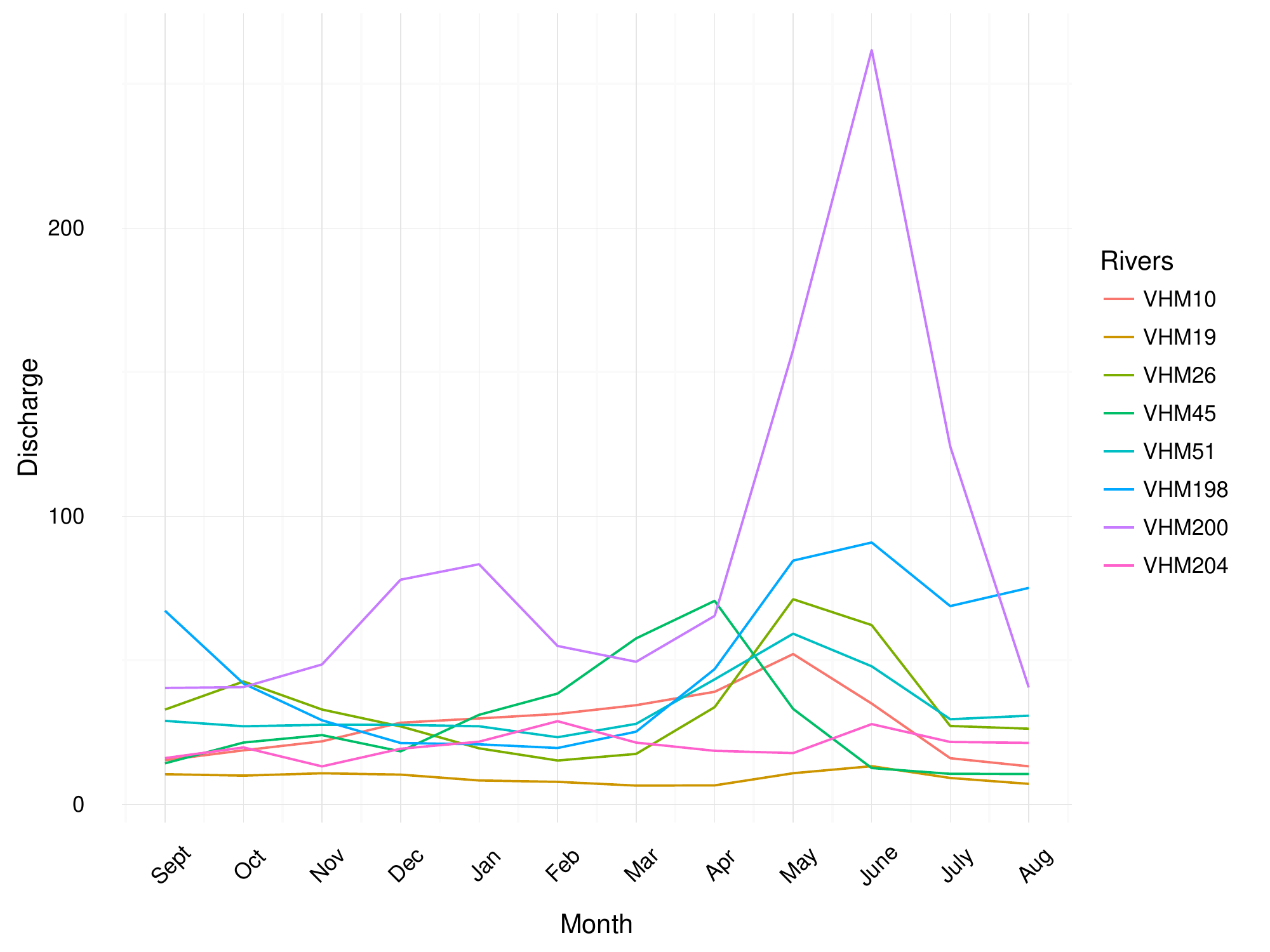}
\caption{Sample means of maximum monthly flow  in $m^3/s$ for each river.} 
\label{fig:meanplot}
\end{figure}

Figure~\ref{fig:meanplot} shows the sample mean of the maximum monthly
instantaneous flow for each river. The catchments have a seasonal
behavior characterised by lower discharge during winter and higher
discharge during spring/summer. The high discharge during spring/summer is mainly due to rising temperatures and snow melt, but the specific timing of the snow melt period varies somewhat for these catchments. 

\subsection{Topographical and Climatic Covariates}

For each catchment, the following topographic and climatic covariates were considered for extrapolating to ungauged catchments: 
\begin{description}
\item{\textbf{Catchment area:}} The area of the river catchment in $\text{km}^2$. 
\item{\textbf{Average precipitation:}} The averaged monthly precipitation over the entire catchment. To construct this covariate the precipitation on a 1 km by 1 km grid over the whole of Iceland was obtained~\citep{crochet2007}, which was then integrated over the catchment area. Finally, the average over all years was found within each month.
\item{\textbf{Maximum daily precipitation:}} Daily precipitation over
  the catchment area within each month was acquired using the same
  method as for the average precipitation. The value corresponding to
  the day with the highest precipitation, cumulated over the
  catchment, was chosen, then the average over all years was found within
  each month.
\item{\textbf{Accumulated precipitation:}} The accumulated precipitation over the catchment since the start of the hydrological year (September). This covariate was potentially useful for explaining high discharge attributed to snow melt.  
\item{\textbf{Average positive temperature:}} Temperature is available on the same grid as precipitation. These values were obtained in the same manner as the average precipitation within each month, with negative values truncated to zero.
\item{\textbf{Maximum positive temperature:}} These values were
  calculated in a similar way to the maximum precipitation values, 
with the difference being that negative temperature values were truncated to zero.
\end{description}

%\newpage
\section{Models and Inference}
\label{sec:model}

\subsection{Preliminary modeling and analysis}
\label{sec:prel-model}

\begin{figure}[h]
%\subfigure[Bandwidth 0.01]{%
%\includegraphics[width=0.5\linewidth]{Figures/P-values/p_values2.pdf}
%\label{fig:psi0}
%}
%\quad
%\subfigure[Bandwidth 0.005]{%
\centering
\includegraphics[width=.75\linewidth]{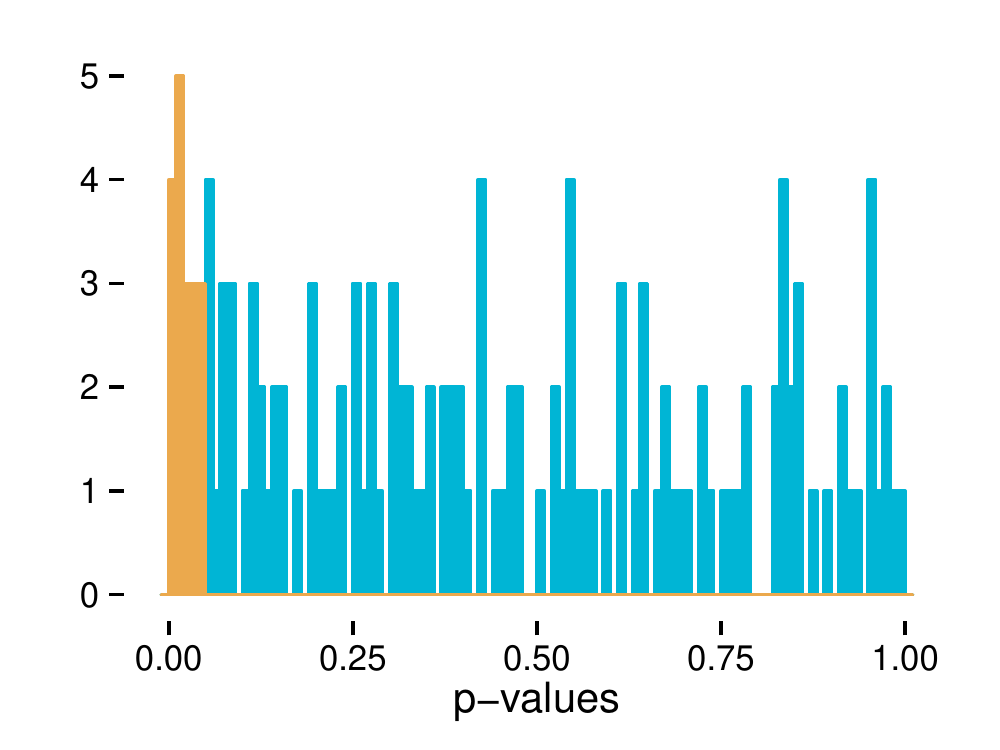}

%}

\caption{A histogram of $p$-values from a Anderson-Darling goodness of fit test for the Gumbel distribution. }
\label{fig:pvalues}

\end{figure}

The Gumbel distribution is a common choice for extreme value data, due to its theoretical foundations~\citep{coles2001}. 
We performed an Anderson--Darling goodness-of-fit test for the Gumbel distribution, for each river and month. The resulting $p$-values are shown in Figure \ref{fig:pvalues}. The empirical distribution of the $p$-values is close to standard uniform, which suggests that the Gumbel distribution fits the observed data reasonably well.

We performed a preliminary analysis of the statistical relationship between maximum instantaneous flow and the topographical and meteorological factors described in Section \ref{sec:data}. The preliminary analysis was carried out as follows. First, maximum likelihood (ML) estimates for both the location and scale parameters of the Gumbel distribution were obtained at all $J=8$ rivers and every month $m$. 
We then fitted log-linear models where the ML estimates of the location and scale parameters, respectively, acted as the response, and all combinations of the aforementioned covariates are assessed. This preliminary analysis revealed a strongly significant log-linear relationship between the ML estimates of the location parameter and catchment area, average precipitation, maximum precipitation and accumulated precipitation. The analysis further showed a strong multicollinearity between average precipitation, maximum precipitation and accumulated precipitation. However, non-significant log-linear relationships were observed between the ML estimates and both average and maximum positive temperature. Based on these results and by using a step-wise log-linear model selection algorithm based on AIC score, it was decided to include both catchment area ($x_1$) and maximum daily precipitation ($x_2$) as predictive covariates for location parameters. Analogous results also hold for the scale parameter.

%
%between the ML estimates of the location and scale parameters and the aforementioned topographical and climate covariates in Section are explored. That is, log-linear models where the ML estimates act as the response and all combinations of the aforementioned covariates are assessed. We select the log-linear model with the strongest AIC score, with a stepwise algorithm [ref]. The preliminary statistical analysis revealed that a log-linear model using catchment area ($x_1$) and maximum daily precipitation ($x_2$) has the strongest AIC score.

\subsection{Description of the proposed hierarchical model}
\label{sec:full-hier-model}

In this section, we present the proposed three-level Bayesian hierarchical model. At the data level, the observed maxima of instantaneous flow $y_{jm,t}$ for river $j$, month $m$, and year $t$ is assumed to follow a Gumbel distribution:
\begin{equation}
  \label{eq:datalevel}
  y_{jm,t} \sim \gumbel(\mu_{jm},\sigma_{jm}),\quad j=1,...,J, \,\, m=1,...,M, \,\, t=1,...,T_{jm}
\end{equation}
where $\mu_{jm}$ and $\sigma_{jm}$ are the location and scale
parameters, respectively. As seen 
from equation~\eqref{eq:datalevel}, these parameters are allowed to differ between both months and rivers. 
At the latent level, the logarithm of the parameters $\mu_{jm}$ and
$\sigma_{jm}$ are modeled with a linear regression model within each month,
incorporating meteorological and topographical covariates. This
approach is inspired by the index flood method, where a linear model
is specified for the logarithm of the mean yearly flow maxima, and is
similar to the model for yearly maxima of
\cite{cunnane1974bayesian}. We build seasonal dependence into the
model, letting latent parameters in neighboring months be \emph{a
  priori} positively correlated. Full details of the model are given below. 

Let $\eta_{jm} = \log \mu_{jm}$ and $\tau_{jm} = \log \sigma_{jm}$.
 The linear model for $\eta_{jm}$ is given by
\begin{equation}
  \label{eq:latentlinmodel}
  \eta_{jm} = (\beta_0 + \beta_{0,m}^*)x_{0,jm} + (\beta_1 + \beta_{1,m}^*)x_{1,jm} + \cdots + (\beta_p + \beta_{p,m}^*)x_{p,jm} + \epsilon_{jm}
\end{equation}
where the $x_{k,jm}$'s are centered log covariates (except $x_{0,jm}=1$ for all $j$ and $m$) and the random effect terms 
$\beta_{k,m}^*$ are given a prior enforcing seasonal behavior, described below. 
This model can be written in matrix form, as follows. Collect the
covariates in the matrix $\bm{X}$, such that the first $M$ rows of
$\bm{X}$ contain the covariates for river $j=1$
over each of the $M$ months,
the next $M$ rows contain  the covariates for river $j=2$, and so
on. Let $(\bm{X}_0, \bm{X}_1, \ldots, \bm{X}_p)$ denote the columns of $\bm{X}$, and let
$$
\bm Z_k = {\rm diag}(\bm{X}_k)(\bm{1}_J \otimes \bm{I}_M )
$$
and
$$
\bm Z = (\bm Z_0, \bm Z_1, \cdots, \bm Z_p).
$$
Further, $\bm \beta = (\beta_0, \beta_1, \ldots, \beta_p)'$, and  let $\bm \eta$ ,
$\bm \beta^*$ and $\bm\epsilon_{\eta}$ contain the $\eta_{jm}$,
$\beta^*_{k,m}$ and $\epsilon_{jm}$, ordered
such that they line up with $\bm X$ and $\bm Z$. Then we may write
$$
\bm\eta = \bm X \bm\beta + \bm Z \bm\beta^* + \bm\epsilon_{\eta}.
$$
The model for $\tau_{jm}$ is similar, with the same covariates, but
different coefficients $\bm\alpha$ and $\bm\alpha^*$ and error term $\bm\epsilon_{\tau}$, and can be written in matrix form as
$$
\bm\tau = \bm X \bm\alpha + \bm Z \bm\alpha^* + \bm\epsilon_{\tau}.
$$
To obtain a latent Gaussian model we must specify multivariate normal priors for the coefficients 
$\bm\alpha$, $\bm\alpha^*$, $\bm\beta$ and $\bm\beta^*$. For $\bm\alpha$ and $\bm\beta$ we fix
$\bm \mu_\alpha$, $\bm \mu_\beta$, $\bm \Sigma_\alpha$ and $\bm \Sigma_\beta$ and set
$$
\quad \pi(\bm\beta) = {\rm N}(\bm\beta|\bm\mu_\beta,\bm\Sigma_\beta), \
\pi(\bm\alpha) = {\rm N}(\bm\alpha|\bm\mu_\alpha,\bm\Sigma_\alpha),
$$
where the choices of $\bm \mu_\alpha$, $\bm \mu_\beta$, $\bm \Sigma_\alpha$ and $\bm \Sigma_\beta$ are explained in Section~\ref{sec:elic-inform-priors}. 
%%The random effects $\bm\alpha^*$ and $\bm\beta^*$ are given the following prior, encoding seasonal dependence:
Let $\bm\beta_k^* = (\beta_{k,1}^*, \ldots, \beta_{k,M}^*)$, $k=0,
\ldots, p$ be the random intercepts ($k=0$) or slopes ($k=1,\ldots,p$)
of covariate $k$  over the $M$ months, and define $\bm\alpha_k^*$
similarly. 
We assume the following priors for $\bm\alpha^*$ and $\bm\beta^*$, encoding seasonal dependence:
$$
\pi(\bm\beta_k^*) = {\rm N}(\bm\beta_k^*|\bm 0,\psi_k^2 \bm Q^{-1}(\kappa)),
\quad 
\pi(\bm\alpha_k^*) = {\rm N}(\bm\alpha_k^*|\bm 0,\phi_k^2 \bm Q^{-1}(\kappa)), 
$$
where $\phi_k^2$ and $\psi_k^2$ are unknown variance parameters
and $\bm Q(\kappa)$ is an $M \times M$ circular precision matrix that has the vector
$$
s \cdot [1 \,\, f_1(\kappa) \,\, f_2(\kappa) \,\, f_1(\kappa) \,\, 1]
= s \cdot [1 \quad -2(\kappa^2 + 2) \quad \kappa^4 + 4\kappa^2 + 6 \quad -2(\kappa^2 + 2) \quad 1 ]
$$ 
on its diagonal band~\citep{lindgren2011explicit}, where $s$ is a
constant ensuring that the inverse of the precision matrix is a
correlation matrix. We have fixed $\kappa$  to the value $\kappa=1$,
giving the prior correlation of 0.67 between neighboring months, which
seems reasonable based on our prior knowledge. Note that $s$ is a
function of $\kappa$, e.g.~for $\kappa=1$, $s \approx 0.268$.

\subsection{Priors for regression coefficients}
\label{sec:elic-inform-priors}

We here present the priors for the regression coefficients $\bm\alpha$ and $\bm\beta$. 
For each $i$, $\alpha_i$ and $\beta_i$ will be given equal priors, since they enter the model in a similar way.
The priors specified below are written in terms of $\beta_i$.
As explained in Section~\ref{sec:full-hier-model},  $\bm\beta$ should be given a multivariate normal prior.
We will assume that the elements $\beta_i$ are \emph{a priori} independent,  so we need to set independent normal priors for the individual coefficients $\beta_i, \ i=0,\ldots,p$.

We start by considering the coefficient $\beta_1$ corresponding to the logarithm 
of the size of the catchment area. First, note that negative values of $\beta_1$ make little sense, as this corresponds to a larger area giving lower maximum flows than a smaller area, other things being equal. To interpret the effects of varying positive values for $\beta_1$, consider precipitation events (rainy clouds) moving over the area. Each event will have a  smaller spatial extent than the catchment area itself, when the catchment area is large; and a hypothetical increase of the catchment area corresponding to given precipitation event will lead to a smaller fraction of the area being covered  by precipitation.  This gives a ``clustering effect'': smaller catchment areas will have a larger proportion covered by precipitation events than larger catchment areas. Since the value $\beta_1=1$ corresponds to a completely uniform distribution of precipitation (which is physically implausible), this means than $\beta_1$ is highly likely to be less than one. In other words, values  $\beta_2>1$ correspond to an effect of area which increases larger than linearly, which is unrealistic for the abovementioned reasons.  

Based on the above, we believe that the most sensible values for $\beta_1$ are in the interval $(0,1)$. We propose that the normal prior density for $\beta_1$ is such that the probability of negative values is $0.05$ and the probability of values greater than one is $0.05$. These values result in a prior mean of $0.5$ and a prior standard deviation of $0.304$.

Considering the effect of precipitation given a fixed area, a similar line of argument can be given for the parameter $\beta_2$ corresponding to maximum daily precipitation: Higher maximum daily precipitation should result in higher flows, so the parameter should be positive. Also, $\beta_2>1$ is unrealistic for similar reasons as explained above for $\beta_1$: natural clustering effects make super-linear effects of precipitation unlikely.  Accordingly, $\beta_2$ is given the same $N(0.5, 0.304^2)$ prior as $\beta_1$.

Since the data should provide good information for the intercept parameter $\beta_0$, there is less of a need to specify an informative prior here. We have therefore chosen a normal density with mean zero and variance $10^4$ as an uninformative prior for the intercept.

\subsection{Penalised complexity priors for hyperparameters}
\label{sec:pc-priors-regression}

In this section, we describe the selection of priors for the hyperparameters $\bm\psi = (\psi_0, \ldots, \psi_p)$ and $\bm\phi = (\phi_0, \ldots, \phi_p)$. We start by considering priors for $\bm\psi$. Note first that $\bm\psi$ can be regarded as a flexibility parameter: $\bm\psi=\bm 0$ corresponds to a restricted model where we set $\bm\beta^*=\bm 0$, i.e.~the \emph{base model}
$$
\bm\eta = \bm X \bm\beta + \bm\epsilon_{\eta}
$$
without correlated random effects. \cite{simpson2014penalising} provide a useful framework for selecting prior densities for flexibility parameters such as $\bm\psi$: penalised complexity (PC) priors. The ideas behind PC priors are thoroughly described in  \cite{simpson2014penalising}, but we give a short review here. PC priors are constructed based on four underlying principles. The first principle is  Occam's razor: we should prefer the simpler base model unless a more complex model is really needed.  The second principle is using the Kullback-Leibler divergence (KLD) as a measure of complexity~\citep{kullback1951information}, where $\sqrt{2\text{KLD}}$ is used to measure the distance between the base model ($\bm\psi=\bm 0$) and the more complex model corresponding to $\bm\psi>\bm 0$  (the factor $2$ is introduced for convenience, giving simpler mathematical derivations). The third principle is that of constant-rate penalisation, which is natural if there is no additional knowledge suggesting otherwise. This corresponds to an exponential prior on the distance scale $d=\sqrt{2\text{KLD}}$. 
Note that defining the prior on the distance scale implies that PC priors are invariant to reparameterization.
The fourth and final principle is  \emph{user-defined scaling}, i.e.~that the user should use (weak) prior knowledge about the size of the parameter to select the parameter of the exponential distribution. \cite{simpson2014penalising} provide both theoretical results and simulation studies showing the PC priors' good robustness properties and strong frequentist performance.

We shall specify independent priors for each component $\psi_k$ of  $\bm\psi$ and each component $\phi_k$ of $\bm \phi$. 
Note that this entails specifying separate base models for each component. While the ideal approach would be to specify an overall multivariate 
PC prior corresponding to the base model, we view this as beyond the scope of this article. 
It is easy to derive that the PC prior approach results in exponential priors for both the $\psi_k$ and $\sigma_\eta$ in this case, see~\cite{simpson2014penalising} for details, so it only remains to specify the scaling, i.e.~the choices of parameters of the respective exponential distributions.

The parameter $\psi_0$ is the standard deviation of the mean zero monthly intercepts $\beta_{0,m}^*$, representing the monthly deviations from the overall intercept $\beta_0$. Since our model is on a logarithmic scale, the values $\beta_{0,m}^*=-4.61$ and $\beta_{0,m}^*=4.61$ correspond to factors $\exp(-4.61)=0.01$ and $\exp(4.61)=100$, respectively, for $\exp(\beta_{0,m}^*)$. Accordingly, $(-4.61, 4.61)$ should be considered to be a wide 95\% probability interval. The value of $\psi_0$ giving this interval is $\psi_0=2.35$. We take $2.35$ as the 0.95 quantile of the prior for $\psi_0$, giving a mean of $0.784$ and a rate of $1.275$ for the exponential prior for $\psi_0$. A similar argument can be given for $\phi_0$, and we give it the same prior as $\psi_0$.

Since $\psi_1$, $\psi_2$, $\phi_1$ and $\phi_2$ have similar roles in the model, they will given identical, independent, priors. We write in terms of $\psi_1$ below, with the understanding that the three other priors are identical.
 It is convenient to use a tail-area argument to specify the scaling. First, consider the sum of the ``fixed effect'' parameter $\beta_1$ and the ``random effect'' parameter $\beta^*_{1,m}$ for some month $m$. For the reasons described in Section~\ref{sec:elic-inform-priors}, most of the prior mass of this  sum should be between zero and one, but the addition of the random effects term will of course increase the variance, so the masses allowed below zero and above one should be larger than the 5\% used in Section~\ref{sec:elic-inform-priors}. We consider 10\% prior mass below zero (and 10\% above one) for $\beta_1+\beta^*_{1,m}$ to give a relatively large mass outside the interval $(0,1)$. This corresponds to a prior standard deviation of approximate $0.32$ for each $\beta^*_{1,m}$. Since this is a high value, it should be in the upper tail of the prior for $\psi_1$: We thus specify that 99\% of the mass of  $\psi_1$ should be below the value $0.32$, giving a rate of approximately $14.4$ (and a mean of approximately  $0.07$) for the exponential prior for $\psi_1$. 

In lack of prior knowledge suggesting otherwise, we give equal priors to $\sigma_\eta$ and $\sigma_\tau$.
The prior for $\sigma_\eta$ can be specified in a more straightforward manner using a direct tail-area argument: Considering the scale of the problem, it seems highly likely that $\sigma_\eta$   should be less than ten, so we put the 0.99-quantile of the exponential prior at the value ten. The result is a rate of  $0.46$ (and a mean of $2.17$).

\subsection{Posterior inference and computation}
\label{sec:computation}

As latent models were imposed on both the location and scale
parameters of the data density, approximation methods such as the
integrated nested Laplace approximation~\citep{rue2009approximate} were inapplicable in our setting.
Therefore, MCMC methods were necessary to make posterior
inference. However, standard MCMC methods such as single site updating
converged slowly and mixed poorly since many model parameters were
heavily correlated in the posterior. For these reasons, all posterior
inference was carried out by using the more efficient MCMC split
sampler~\citep{geirsson2015mcmc}. The MCMC split sampler is a
two-block Gibbs sampling scheme designed for LGMs, where tailored
Metropolis--Hastings strategies are implemented within in both
blocks. The sampling scheme is well suited to infer LGMs with
non-Gaussian data density where latent models are imposed on both the
location and scale parameters.

The main idea of the MCMC split sampler is to split the latent Gaussian parameters into two vectors, called the ``data-rich'' block and the ``data-poor'' block. The data-rich block consists of the parameters that enter directly into the likelihood function, in our case the location parameters $\mu_{jm}$ and the scale parameters $\sigma_{jm}$, for $j=1,\ldots,J$ and $m=1,\ldots,M$. The data-poor block consists of the remaining parameters (in our case, including the regression parameters and hyperparameters). An efficient block Gibbs sampling scheme can then be implemented by sampling from the full conditional distributions of each block. For the data-poor block, it turns out that the full conditional is multivariate Gaussian, so sampling can be done quickly using a version of the one-block sampler of~\citet{knorr2002block}. The data-rich block can also be sampled efficiently, for details see \citet{geirsson2015mcmc}.

%\newpage
\section{Results}
\label{sec:results}

The model described in Section~\ref{sec:model} was fitted using the
MCMC split sampler, with 30000 iterations, discarding a
burn-in of 10000. Runtime on a modern desktop (Ivy Bridge Intel Core
i7-3770K, 16GB RAM and solid state hard drive), was approximately one hour. All the calculations were done using \texttt{R}. 

\begin{figure}[p]
\subfigure[$\beta_{0}$ (intercept)]{%
\scalebox{0.9}{\includegraphics[width=0.33\linewidth]{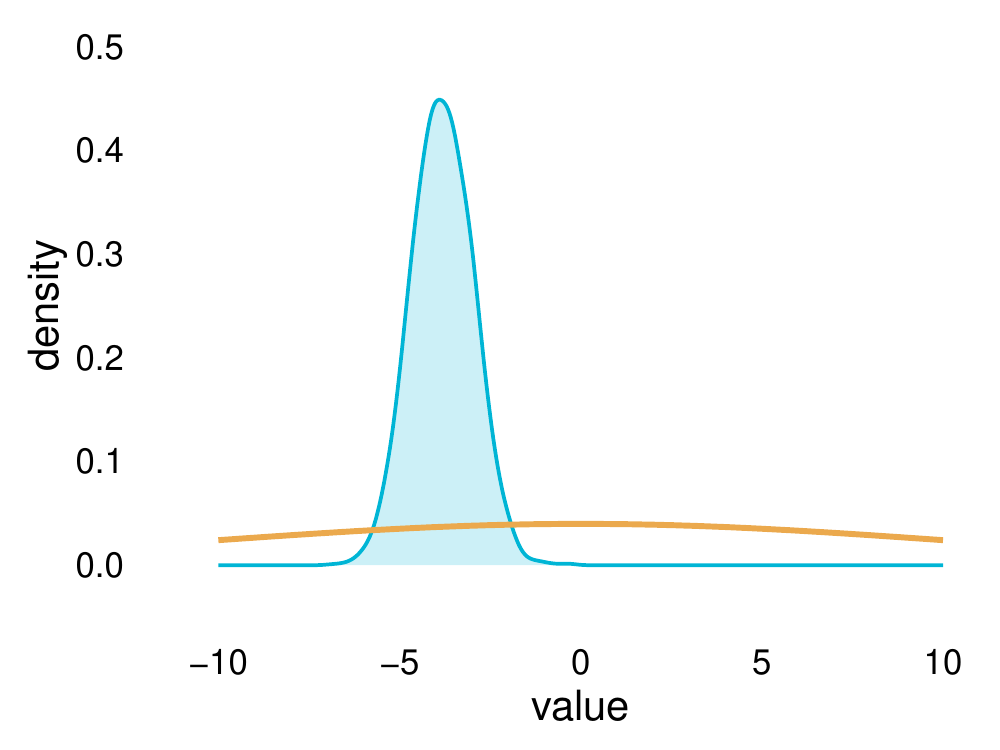}}
\label{fig:beta0}
}
\subfigure[$\beta_{1}$ (catchment area)]{%
\scalebox{0.9}{\includegraphics[width=0.33\linewidth]{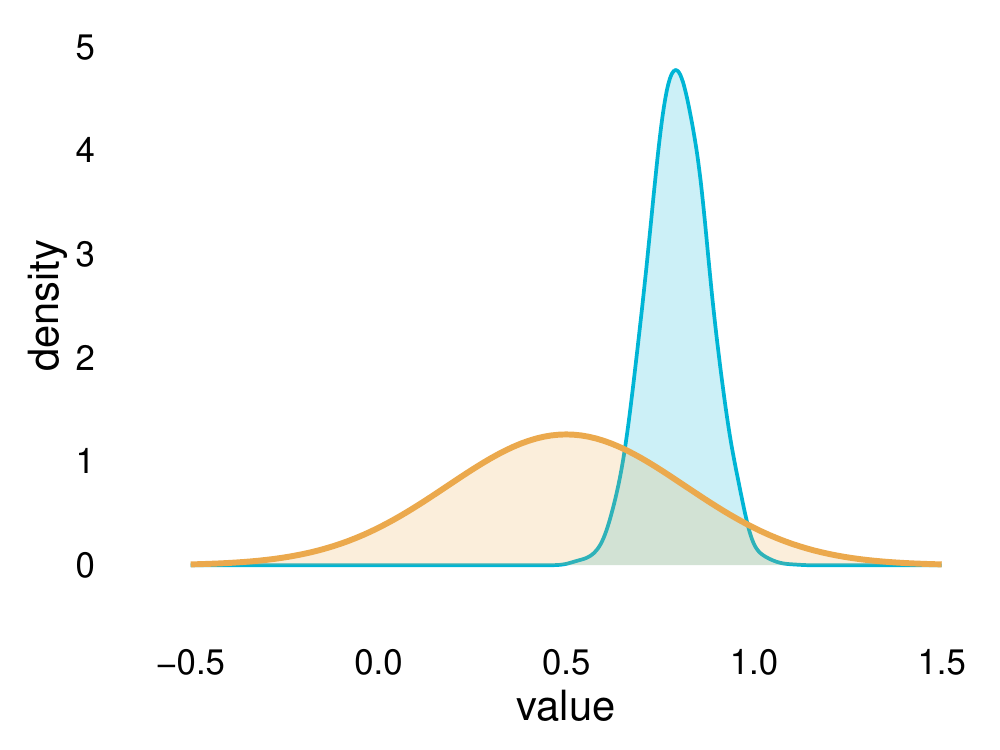}}
\label{fig:beta1}
}
\subfigure[$\beta_{2}$ (maximum precip.)]{%
\scalebox{0.9}{\includegraphics[width=0.33\linewidth]{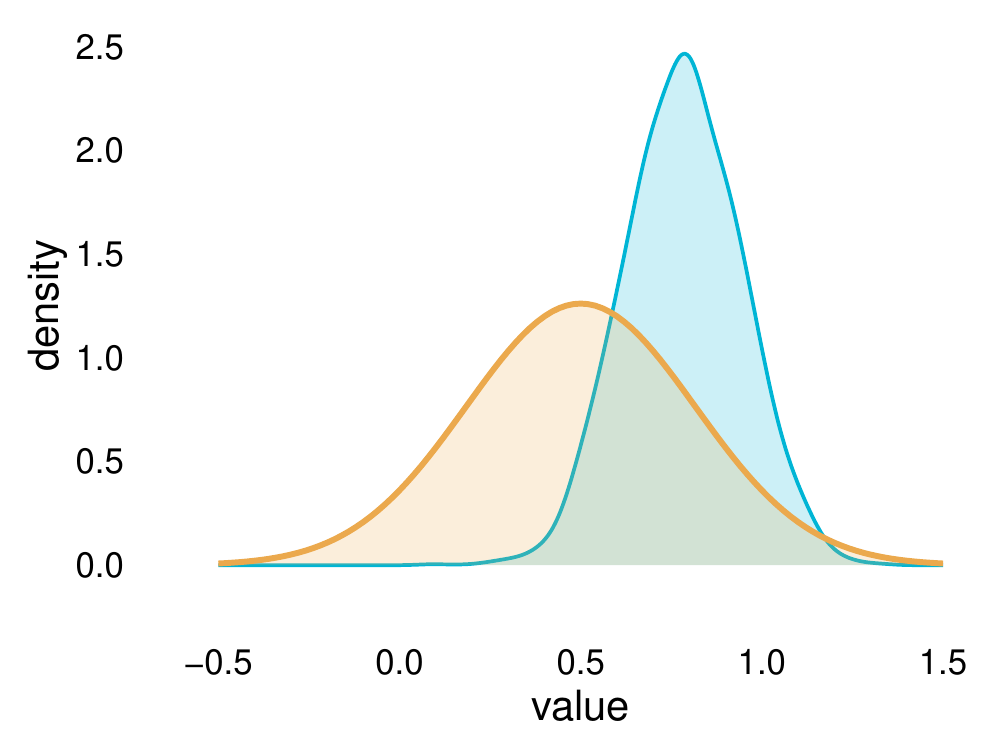}}
\label{fig:beta2}
}

\subfigure[$\alpha_{0}$ (intercept)]{%
\scalebox{0.9}{\includegraphics[width=0.33\linewidth]{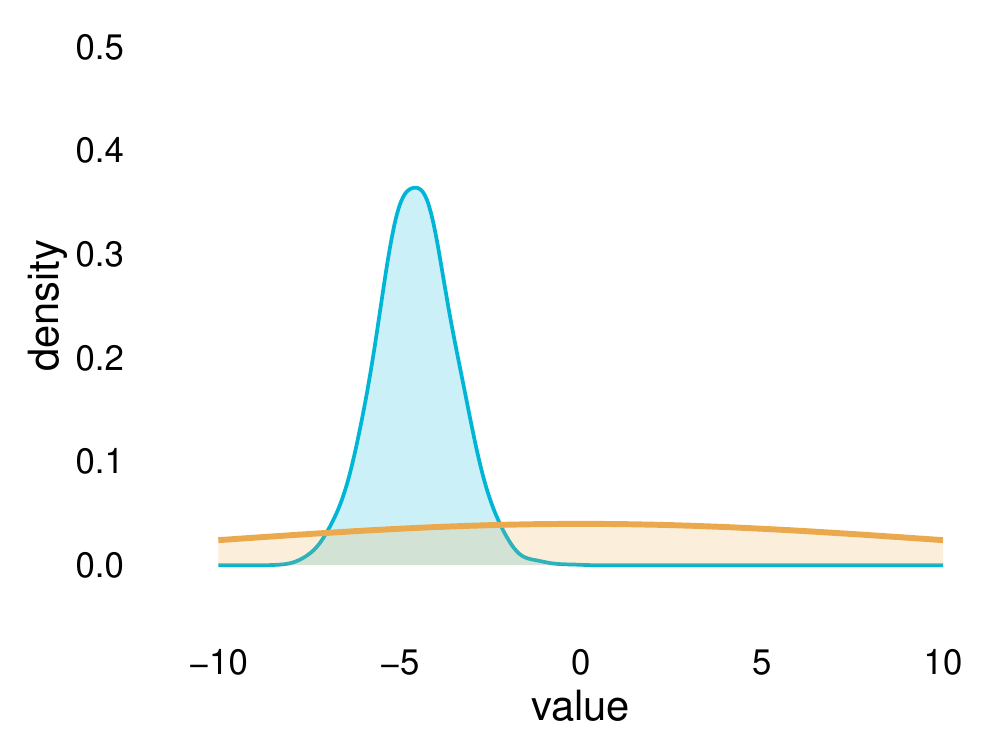}}
\label{fig:alpha0}
}
\subfigure[$\alpha_{1}$ (catchment area)]{%
\scalebox{0.9}{\includegraphics[width=0.33\linewidth]{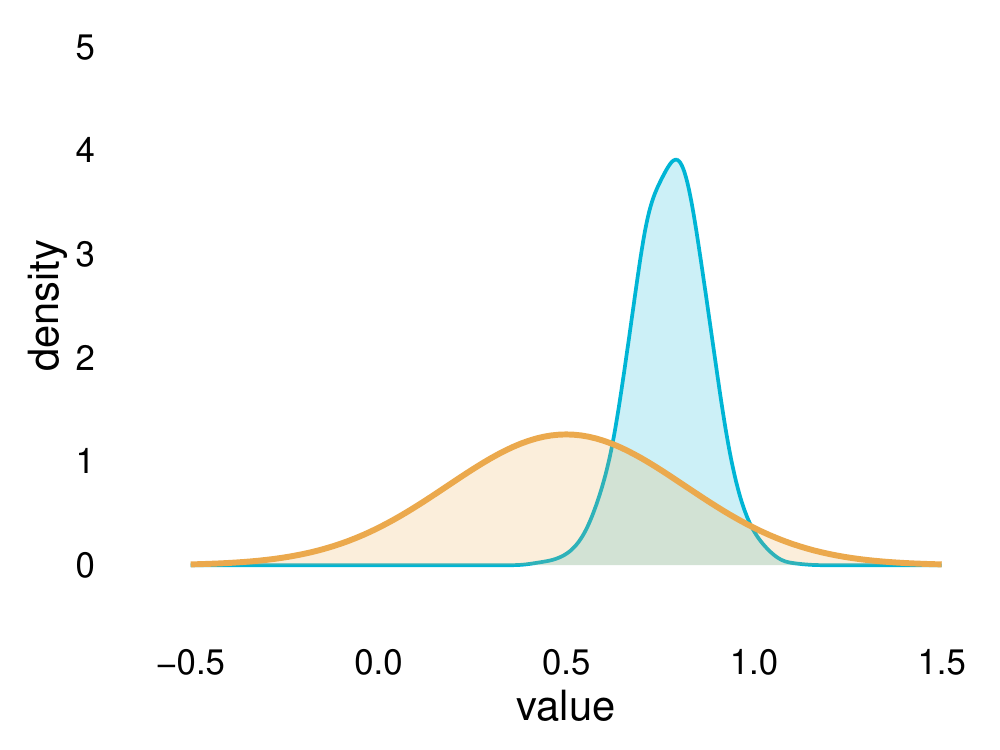}}
\label{fig:alpha1}
}
\subfigure[$\alpha_{2}$  (maximum precip.)]{%
\scalebox{0.9}{\includegraphics[width=0.33\linewidth]{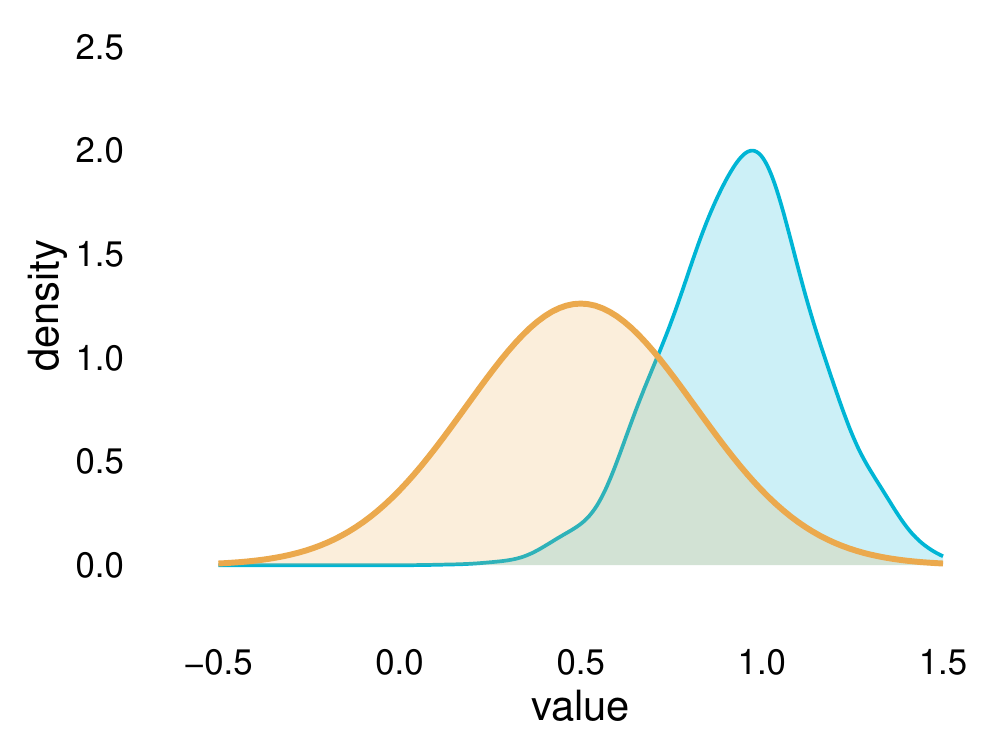}}
\label{fig:alpha2}
}
\caption{Prior (orange) and posterior densities (blue) for the regression coefficients. }
\label{fig:PriorVsPost_regression}

\end{figure}

Figure \ref{fig:PriorVsPost_regression} shows prior densities (in
orange) together with posterior densities (light
blue) for the regression coefficients $\bm \beta$ and $\bm \alpha$. The
posteriors look close to being normally distributed. We
see that the intercepts (Figures \ref{fig:beta0} and
\ref{fig:alpha0}) are well identified, with modes close to $-5$ and
$-4$, respectively, even though they have a
vague prior. This is as expected, since the intercepts correspond
to an overall, ``average'' level which should be relatively easy to
infer. The posteriors for the regression coefficients $\beta_1$ and
$\alpha_1$, corresponding to
log catchment area, (Figures \ref{fig:beta1} and \ref{fig:alpha1}),
look similar, though the posterior for $\alpha_1$ (in the model for
log scale) is slightly wider. Both have a mode of around 0.75, 
and most of the posterior mass in the region between 0.5
and 1.
 Posteriors
for $\beta_2$ and $\alpha_2$, corresponding to  maximum daily
precipitation  (Figures \ref{fig:beta2} and \ref{fig:alpha2}) are
wider than those for $\beta_1$ and $\alpha_1$, with most of
the mass in the region between 0.4 and 1.5.  The posterior mode of
$\beta_2$ is around 0.9, while the posterior mode of $\alpha_2$ is close to
1.0.

%%%%
%Covariate coeffiecients
%%%%

\begin{figure}[p]
\subfigure[$\psi_{0}$ (intercept)]{%
\includegraphics[width=0.32\linewidth]{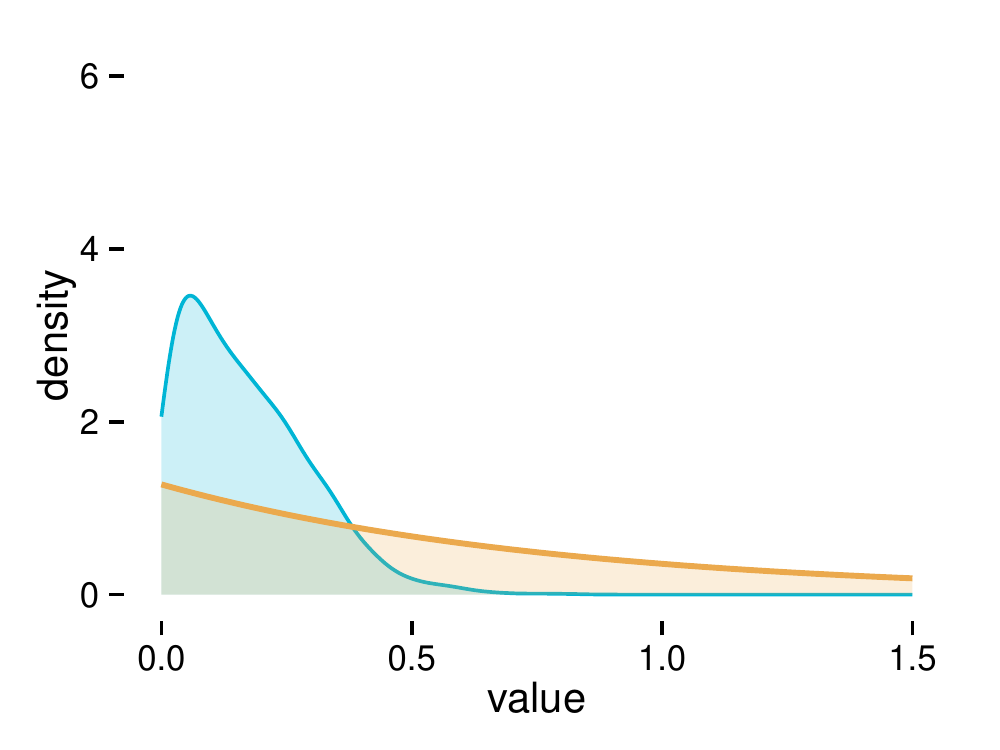}
\label{fig:psi0}
}
\subfigure[$\psi_{1}$ (catchment area)]{%
\includegraphics[width=0.32\linewidth]{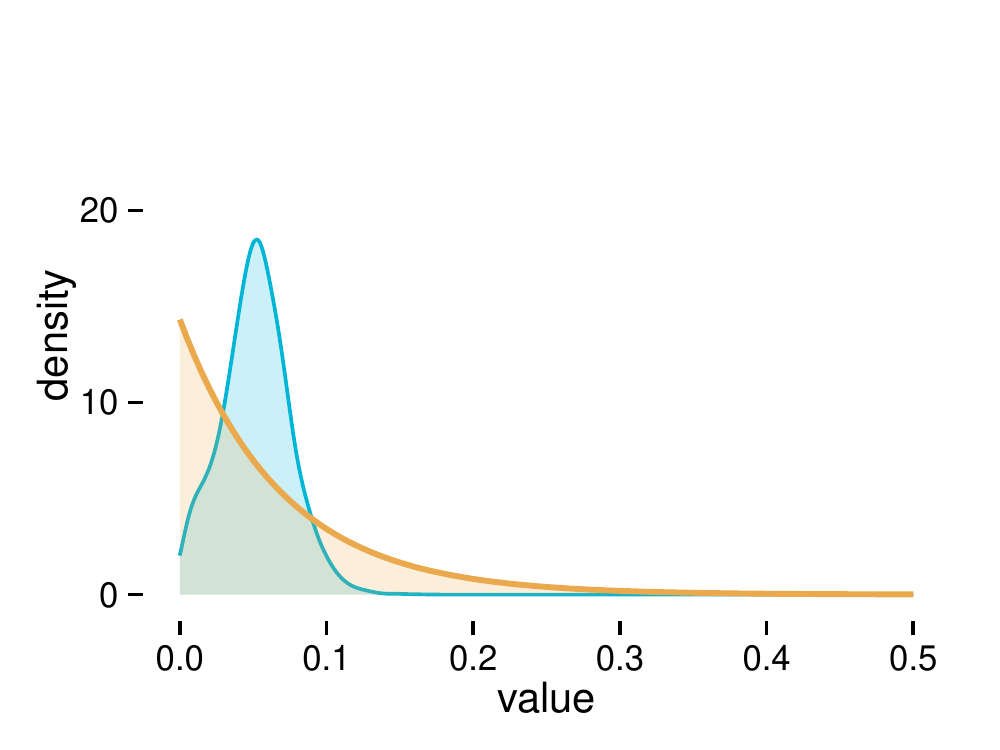}
\label{fig:psi1}
}
\subfigure[$\psi_{2}$ (maximum precip.)]{%
\includegraphics[width=0.32\linewidth]{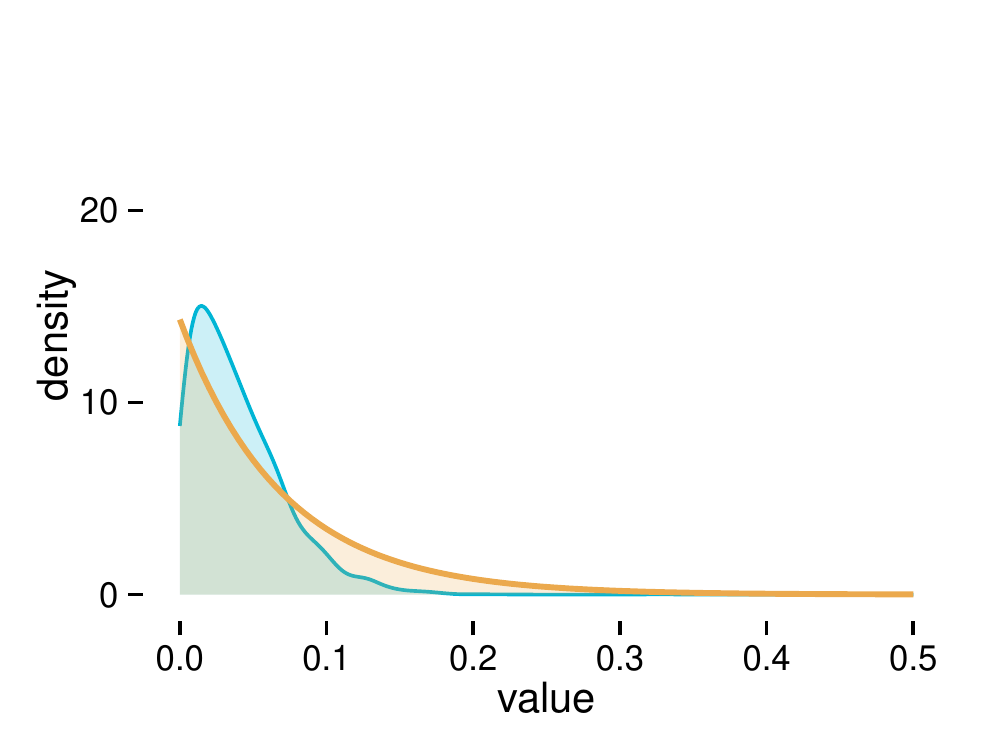}
\label{fig:psi2}
}

\subfigure[$\phi_{0}$ (intercept)]{%
\includegraphics[width=0.32\linewidth]{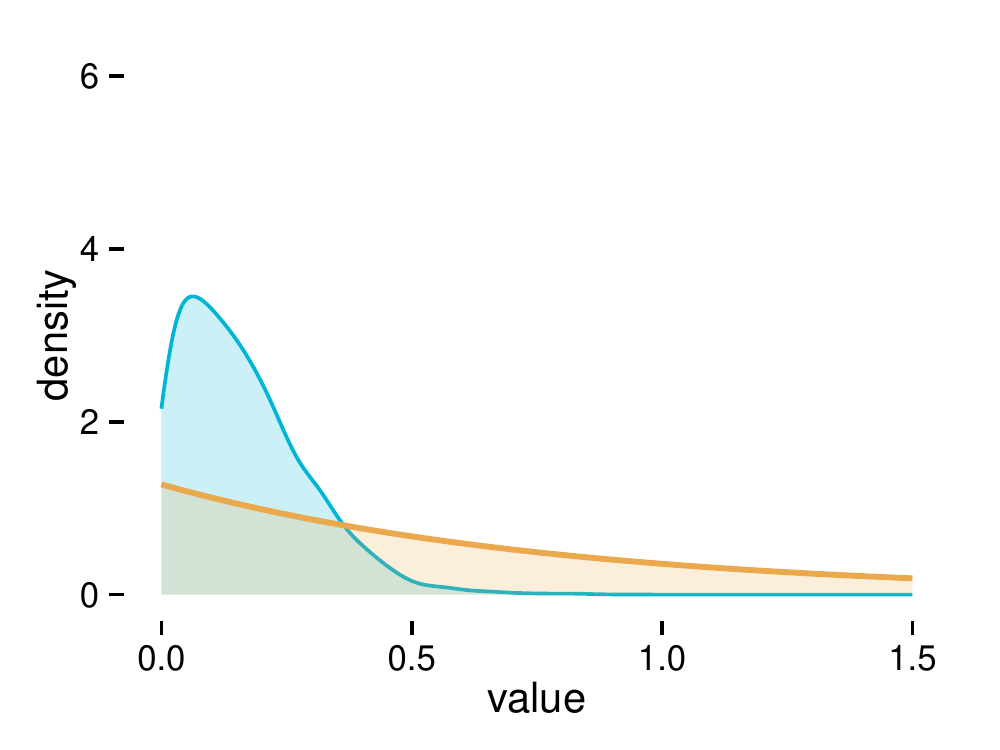}
\label{fig:phi0}
}
\subfigure[$\phi_{1}$ (catchment area)]{%
\includegraphics[width=0.32\linewidth]{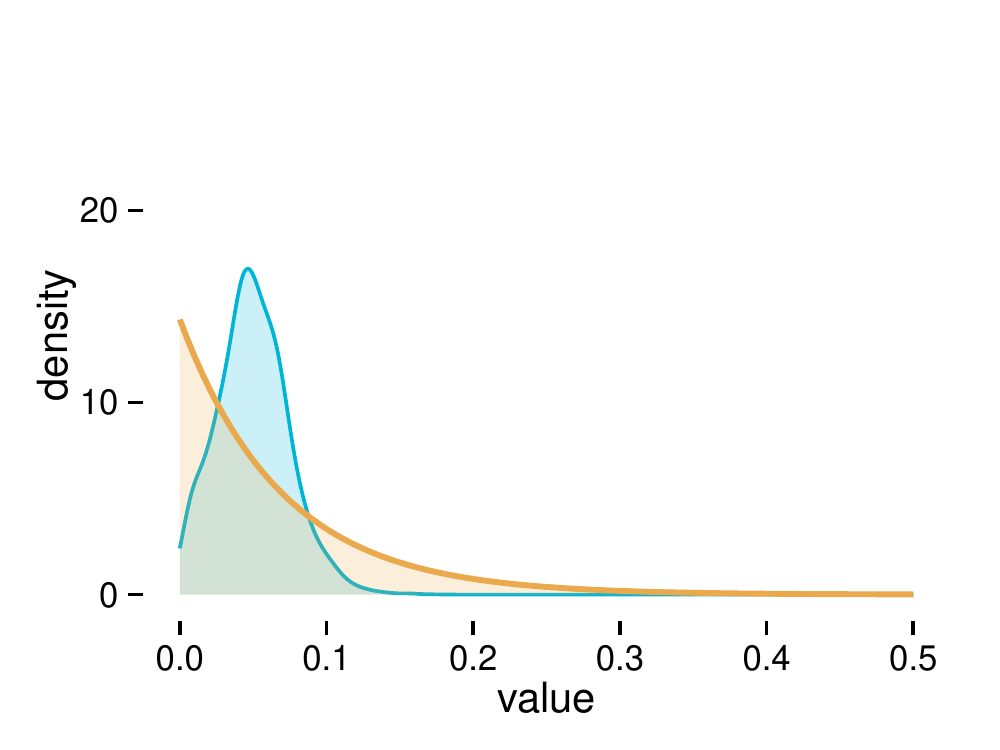}
\label{fig:phi1}
}
\subfigure[$\phi_{2}$ (maximum precip.)]{%
\includegraphics[width=0.32\linewidth]{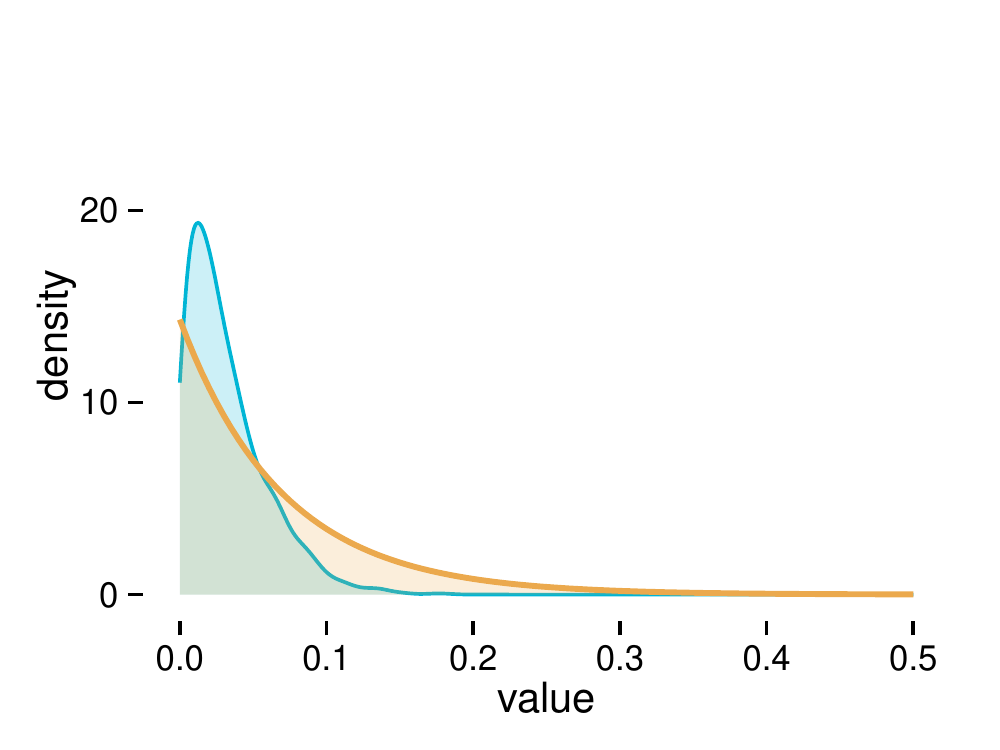}
\label{fig:phi2}
}

\subfigure[$\sigma_\eta$ (error term)]{%
\includegraphics[width=0.32\linewidth]{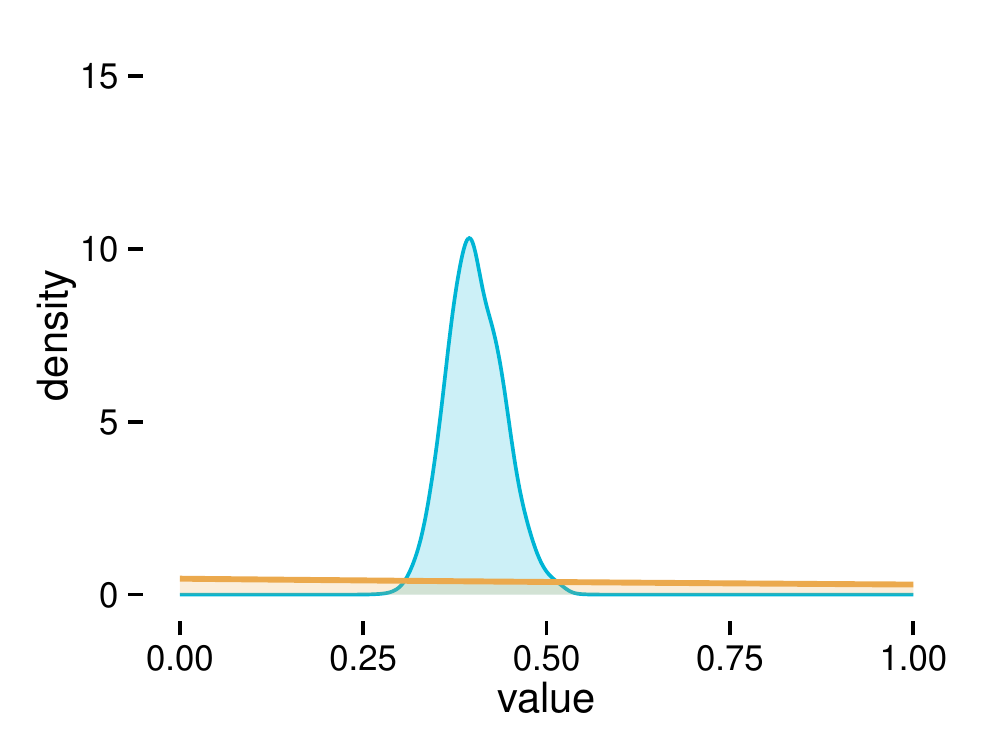}
\label{fig:sigmaeta}
}
\subfigure[$\sigma_\tau$ (error term)]{%
\includegraphics[width=0.32\linewidth]{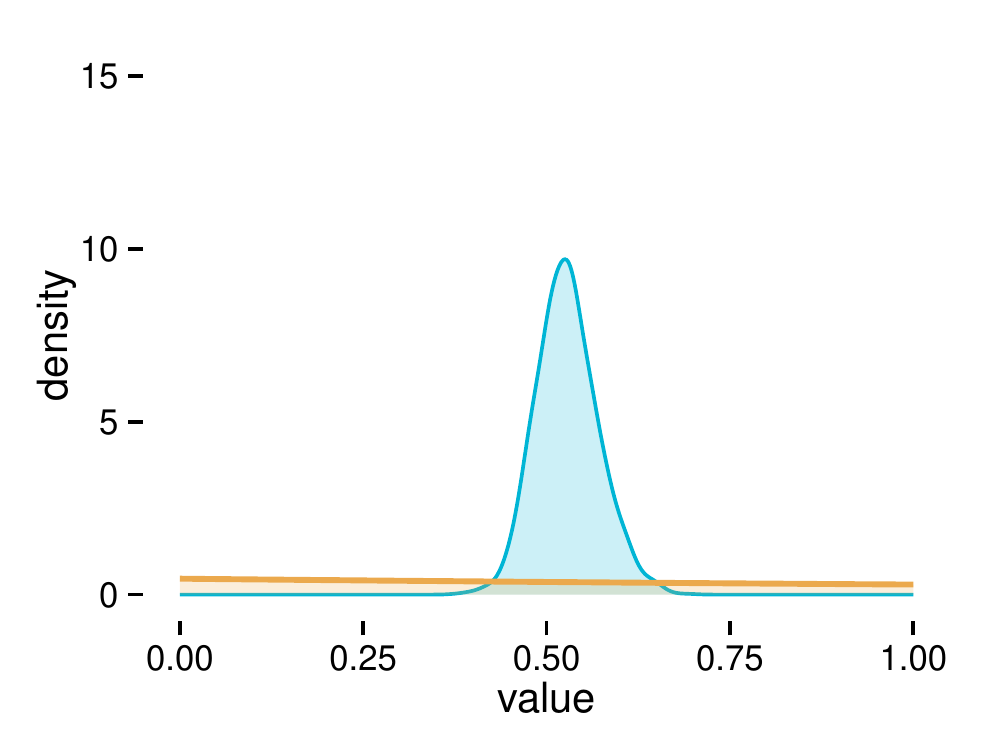}
}
\label{fig:sigmatau}

\caption{Prior (orange) and posterior (blue) densities of the hyperparameters.}

\label{fig:PriorVsPost_hyper}

\end{figure}

Figure~\ref{fig:PriorVsPost_hyper} shows prior and posterior densities for all
eight hyperparameters of the model. We see that the hyperparameters for
the random effects' standard deviations 
$\psi_i$ and $\phi_i$ ($i=0,1,2$) are all shrunk somewhat towards
zero. However, the posterior mode is larger than zero for all
hyperparameters, particularly for $\phi_1$, where there is very little
mass close to zero. For the standard deviations $\phi_1, \phi_2,
\psi_1$ and $\psi_2$
most of the posterior mass is between 0 and 0.1, while $\phi_0$ and
$\psi_0$ (corresponding to the random intercepts) have most of their
posterior mass between 0 and 0.5.
  Posteriors for $\sigma_\eta$
and $\sigma_\tau$ (the two residual noise standard deviations of the
model) are well identified, even though they were given an very weakly
informative prior. The posterior modes of  $\sigma_\eta$
and $\sigma_\tau$ are close to 0.5.

\begin{figure}[p]
\subfigure[$\bm \beta_0^*$ (intercept)]{%
\includegraphics[width=0.32\linewidth]{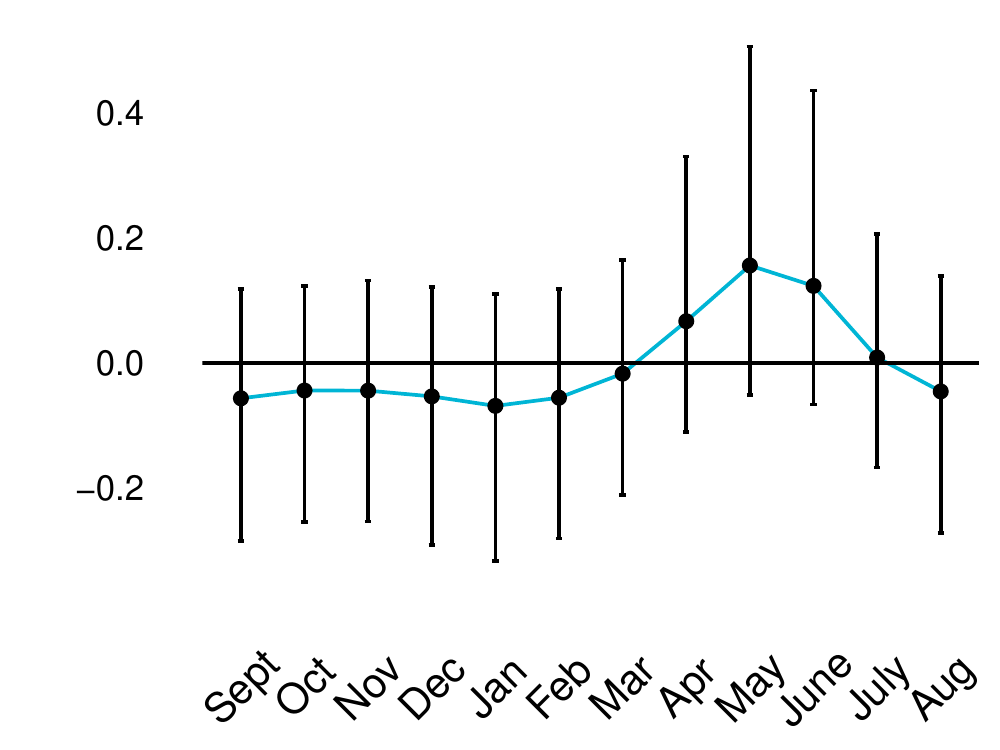}
\label{fig:betastar_0}
}
\subfigure[$\bm \beta_1^*$ (catchment area)]{%
\includegraphics[width=0.32\linewidth]{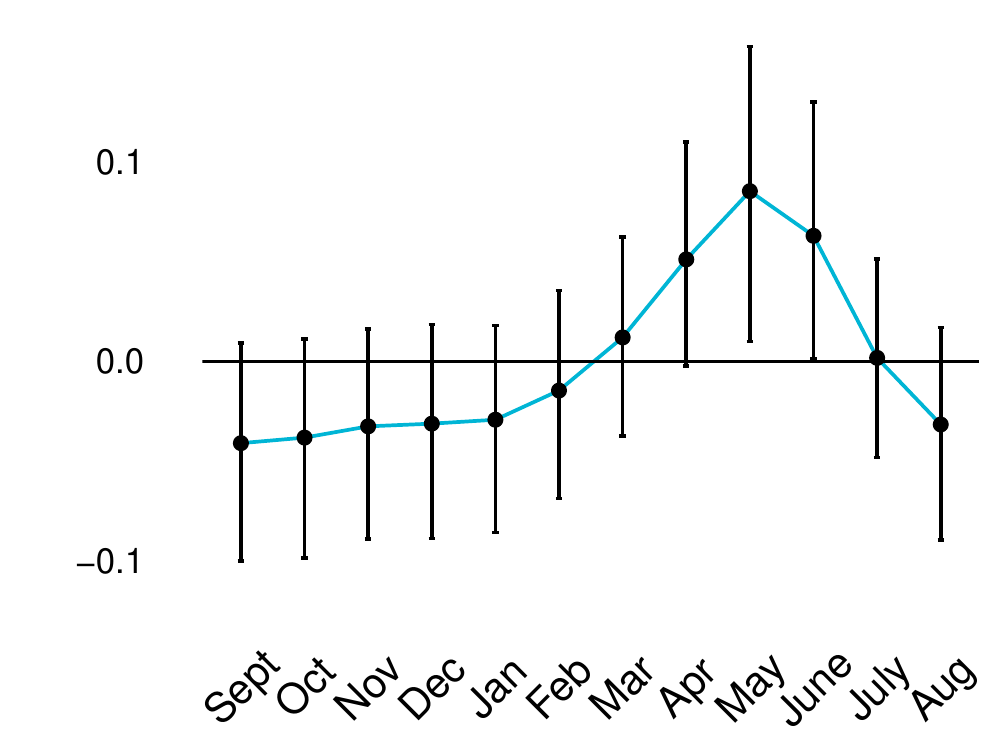}
\label{fig:betastar_1}
}
\subfigure[$\bm \beta_2^*$ (maximum precip.)]{%
\includegraphics[width=0.32\linewidth]{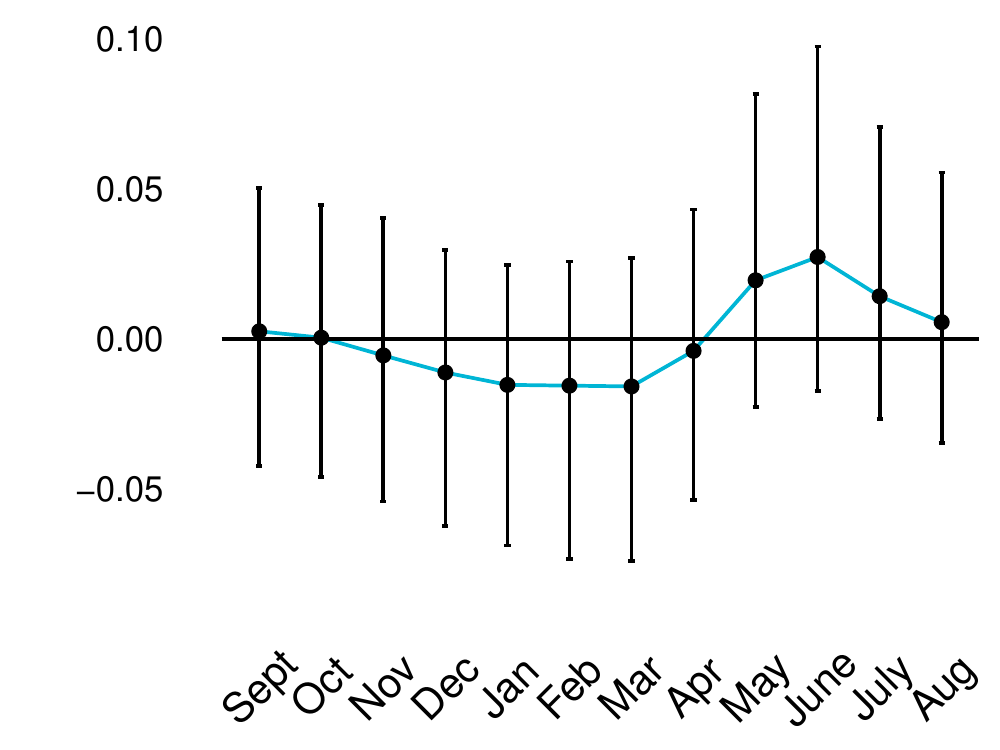}
\label{fig:betastar_2}
}

\subfigure[$\bm \alpha_0^*$ (intercept)]{%
\includegraphics[width=0.32\linewidth]{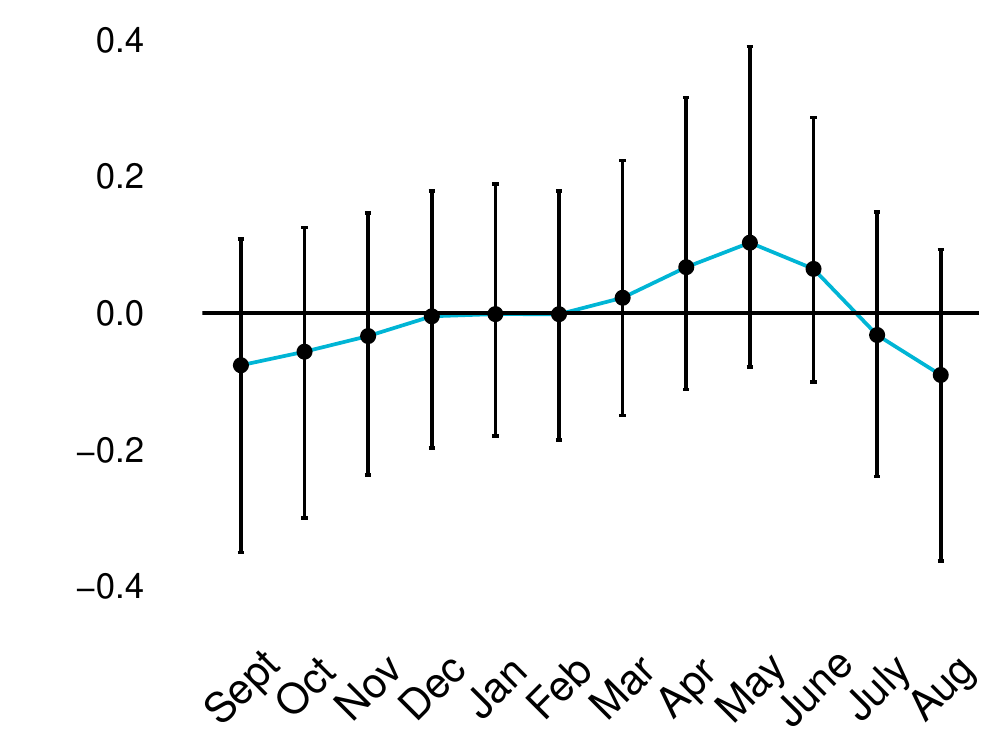}
\label{fig:mustar_0}
}
\subfigure[$\bm \alpha_1^*$ (catchment area)]{%
\includegraphics[width=0.32\linewidth]{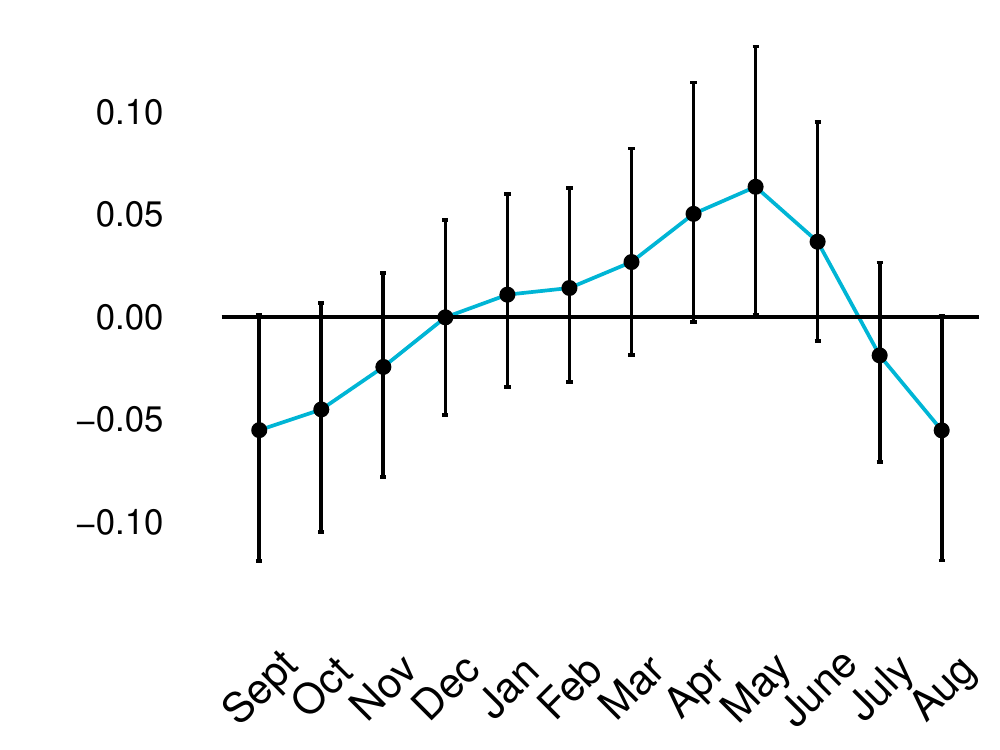}
\label{fig:mustar_1}
}
\subfigure[$\bm \alpha_2^*$ (maximum precip.)]{%
\includegraphics[width=0.32\linewidth]{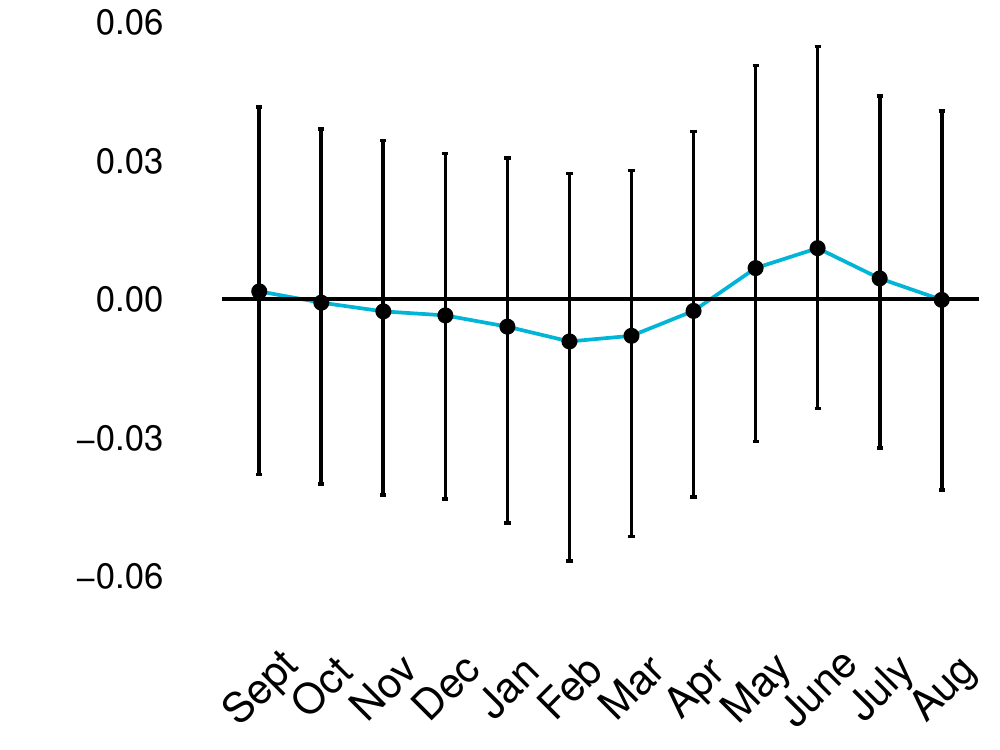}
\label{fig:mustar_2}
}

\caption{Posterior mean and posterior 80\% intervals for the seasonal
  effects.}

\label{fig:seasonal}

\end{figure}

Figure \ref{fig:seasonal} shows the seasonal effects, together with
80\% pointwise credible intervals.
 It
seems like there is some evidence for a seasonal effect for $\bm
\beta_0^*$ (the intercept of the location model), and $\bm \beta_1^*$
and $\bm \alpha_1^*$ (corresponding to catchment area), while this is not so
clear for the other parameters. This is consistent with what was
seen in Figure~\ref{fig:PriorVsPost_hyper}, particularly when
comparing the posterior for $\phi_1$ with the corresponding seasonal
effect for $\bm \alpha_1^*$.

%%%%
%Hyper parameters
%%%%

%%%
%Time effect
%%%

\begin{figure}[p]
\subfigure[VHM10, May]{%
\includegraphics[width=0.32\linewidth]{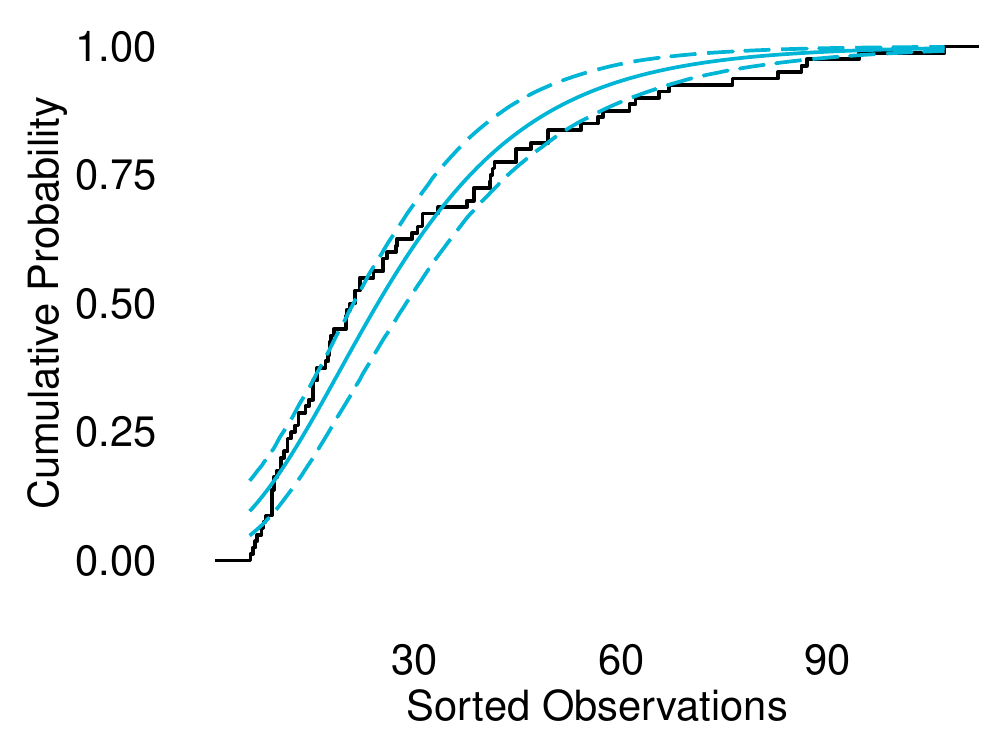}
\label{fig:cdf_1}
}
\subfigure[VHM19, December]{%
\includegraphics[width=0.32\linewidth]{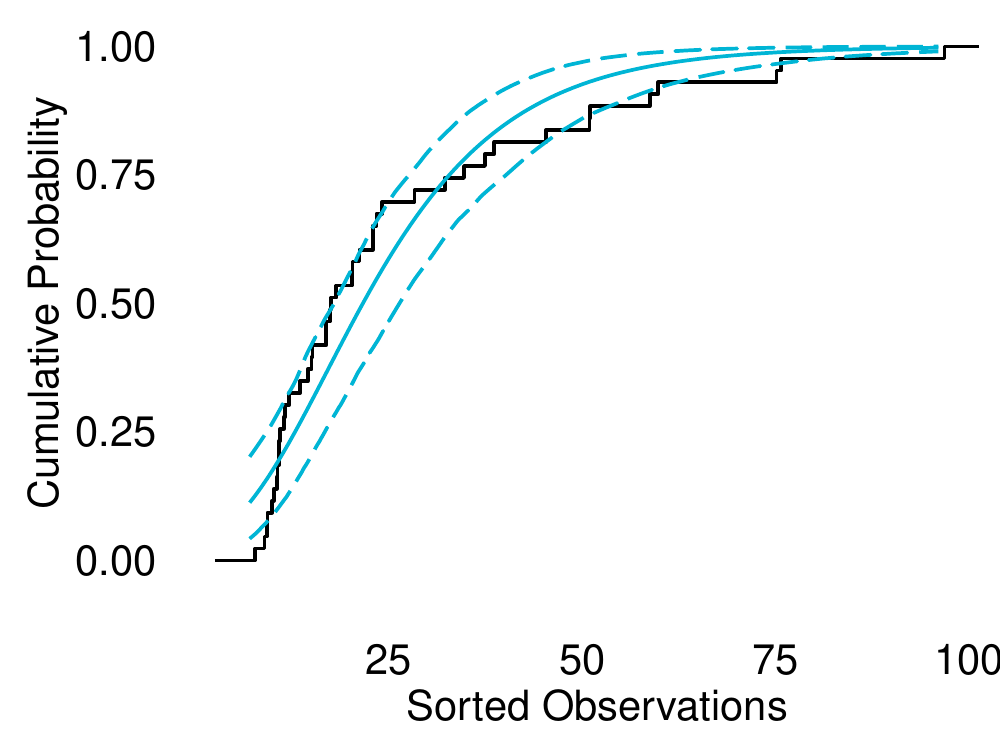}
\label{fig:cdf_35}
}
\subfigure[River VHM45, August]{%
\includegraphics[width=0.32\linewidth]{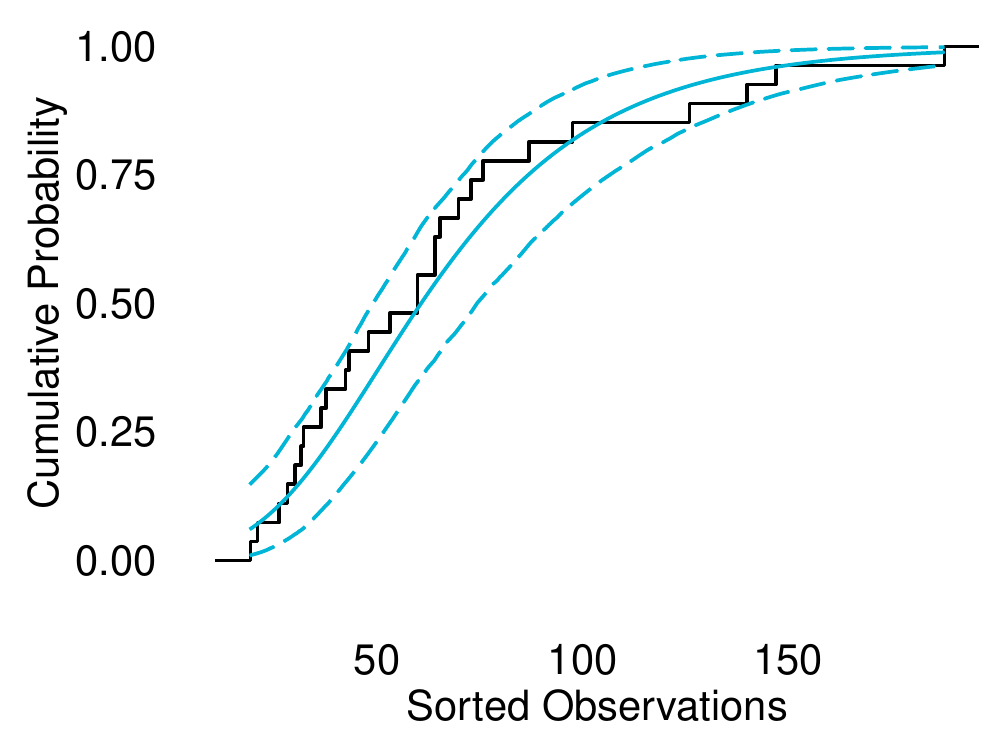}
\label{fig:cdf_101}
}

\subfigure[VHM10, May]{%
\includegraphics[width=0.32\linewidth]{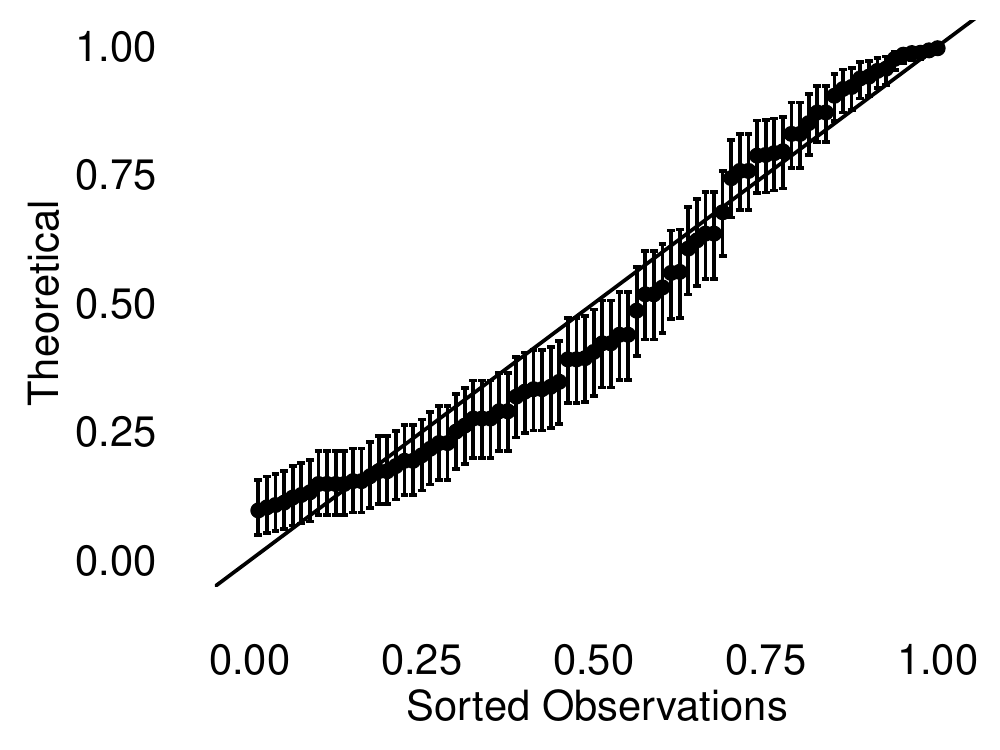}
\label{fig:qq_1}
}
\subfigure[VHM19, December]{%
\includegraphics[width=0.32\linewidth]{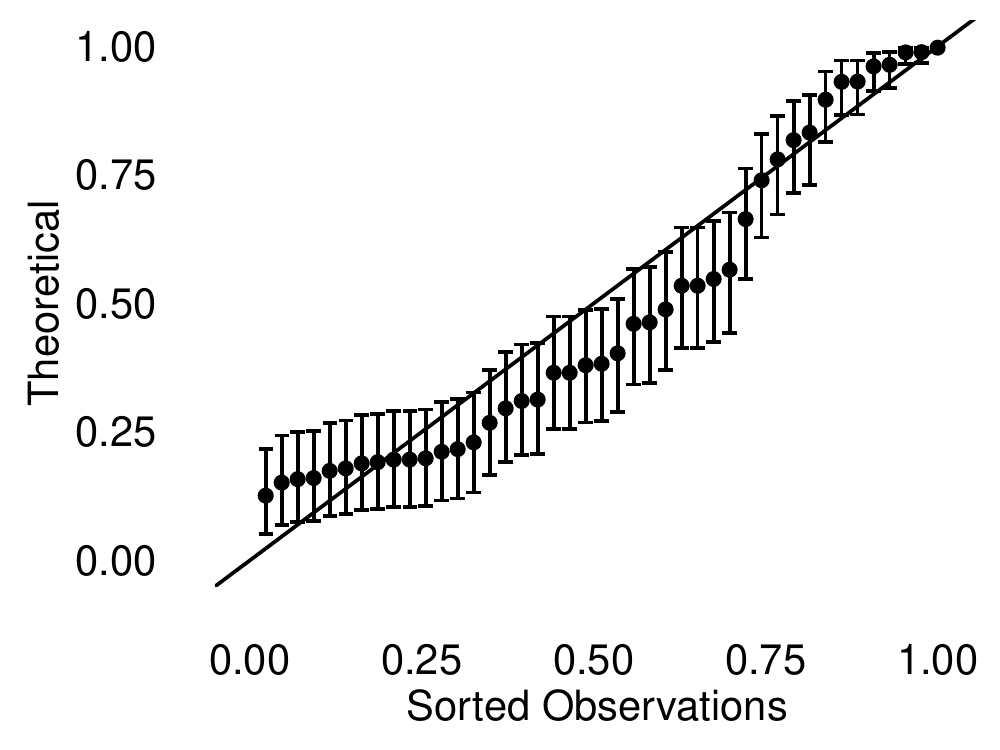}
\label{fig:qq_35}
}
\subfigure[VHM45, August]{%
\includegraphics[width=0.32\linewidth]{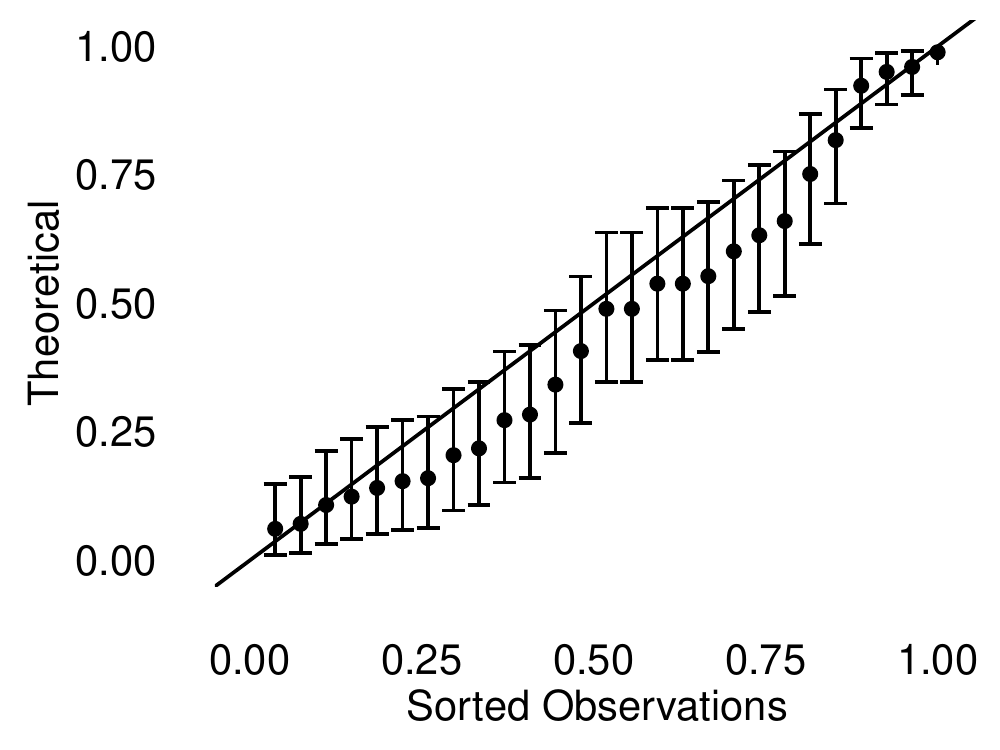}
\label{fig:qq_101}
}
\caption{Model fit. The top panel shows predicted vs empirical
  cumulative distribution functions for three randomly chosen
  river-month combinations. The bottom panel shows
  probability-probability (pp) plots for the same river-month
  combinations, i.e.~the empirical CDF is plotted against the 
CDF predicted from the model.}

\label{fig:cdfqq}

\end{figure}

The left panels of Figure \ref{fig:cdfqq} show empirical cumulative distribution
functions (CDFs) together with CDFs predicted from the model, for
three randomly chosen river/month-combinations. 
The right panels show corresponding
PP plots, i.e.~the empirical CDF is plotted against the 
CDF predicted from the model for each river and each month. Uncertainty bands correspond to
pointwise 95\% credible intervals. The model seems to fit the data reasonably
well.

%%%
%QQ and CDF plots
%%%

\begin{figure}[!h]
\centering
\subfigure[VHM10]{%
\includegraphics[scale=0.39]{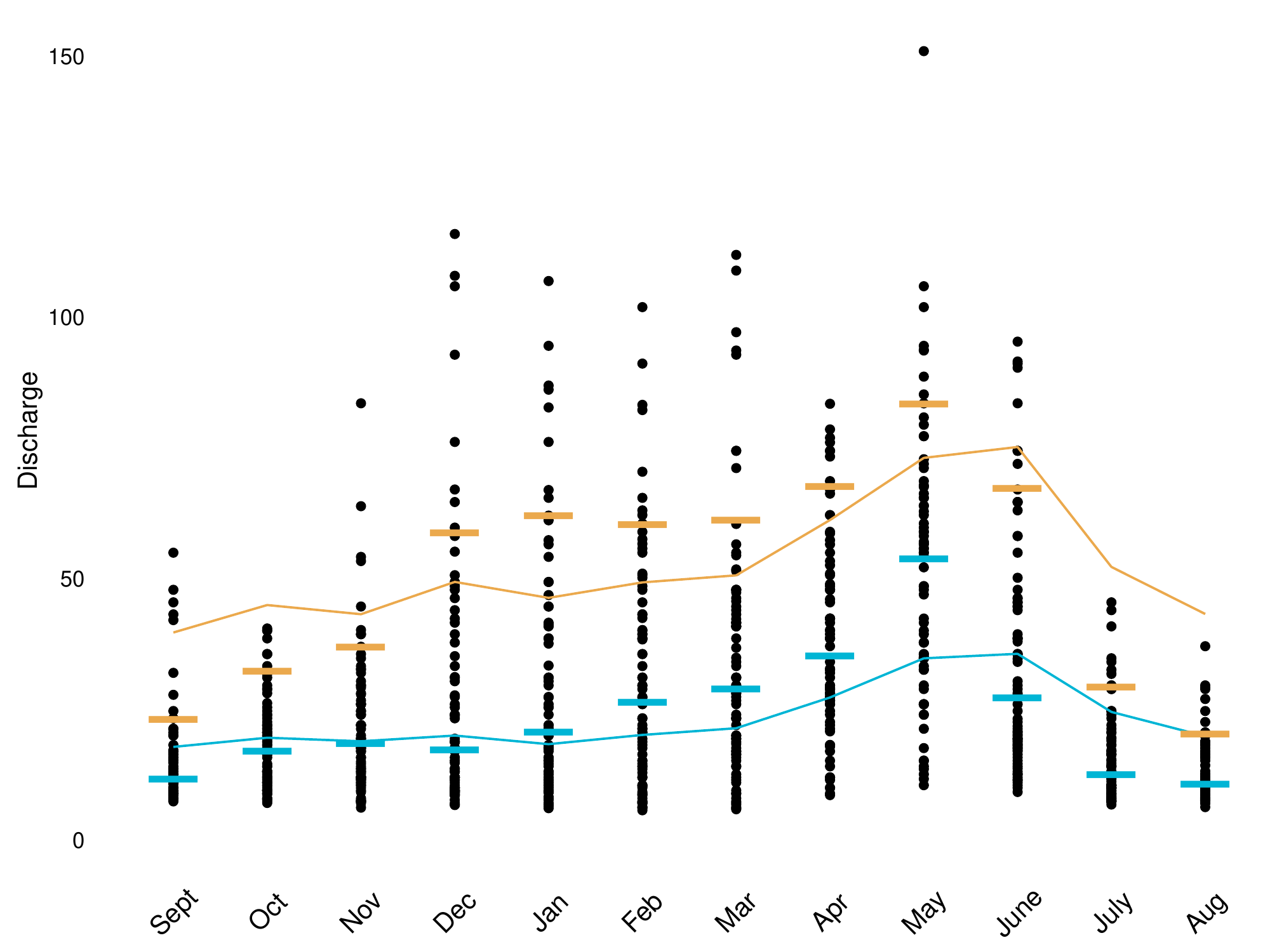}
\label{fig:pred1}
}
\subfigure[VHM19]{%
\includegraphics[scale=0.39]{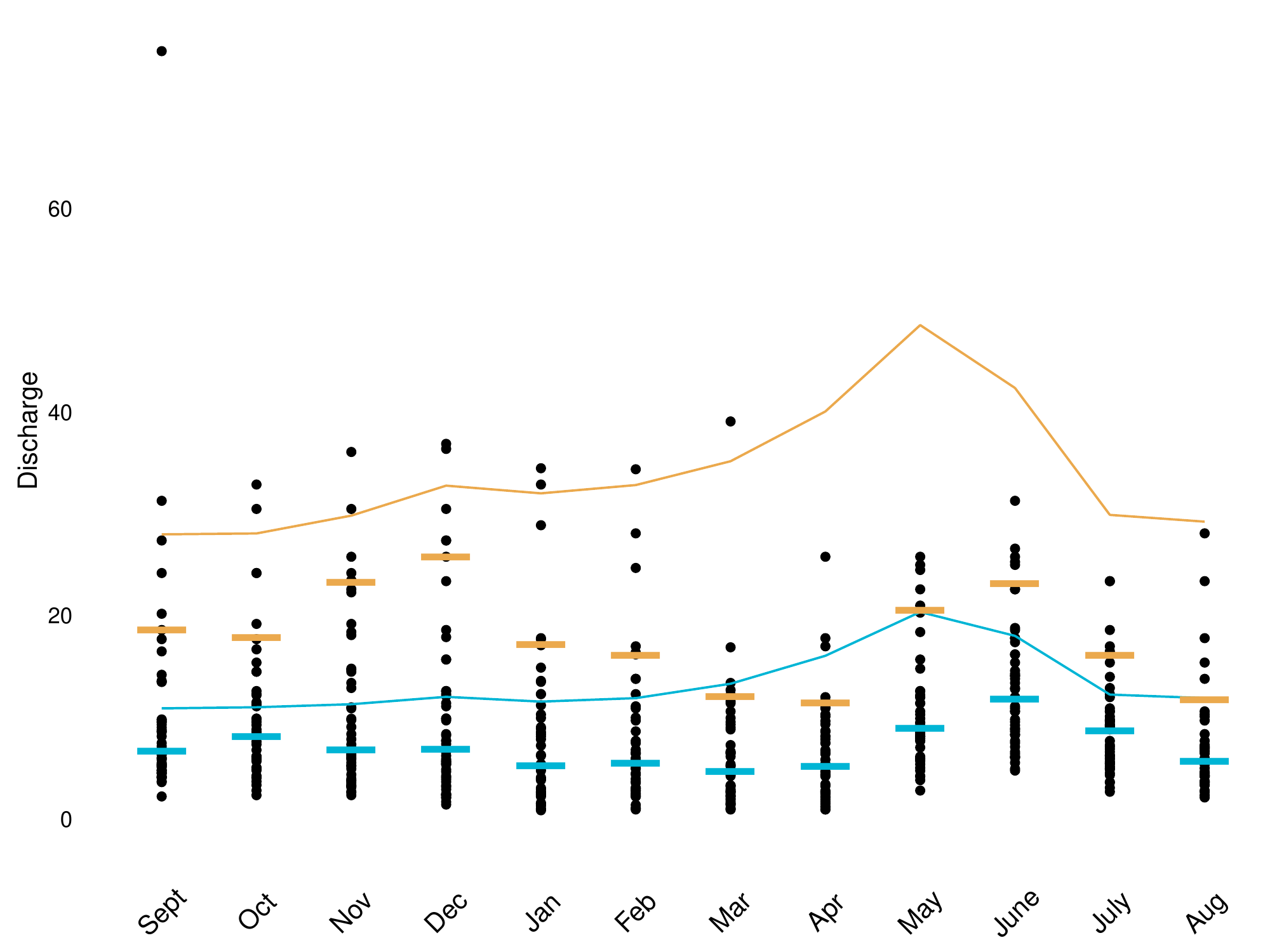}
\label{fig:pred2}
}

\subfigure[VHM26]{%
\includegraphics[scale=0.39]{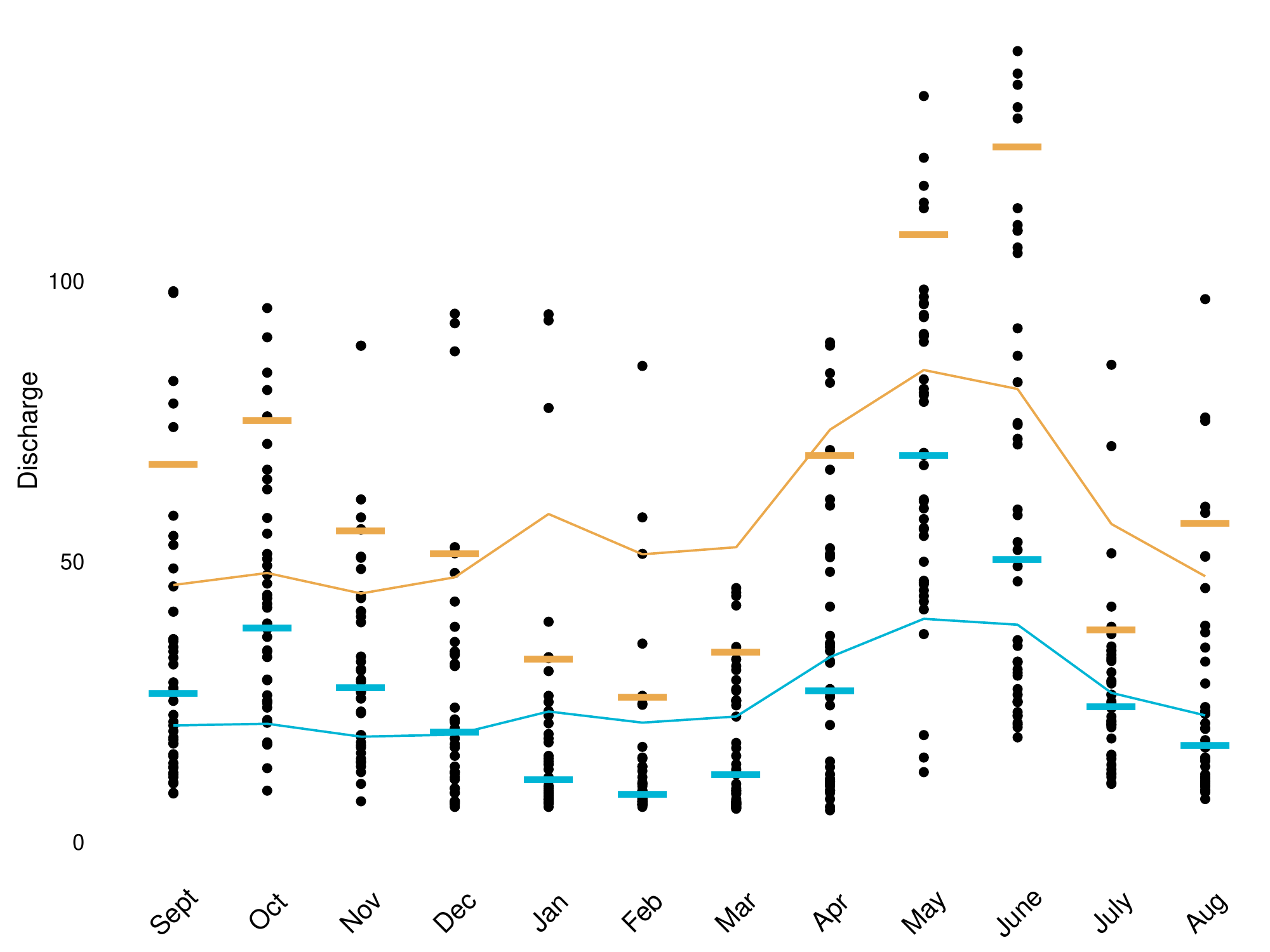}
\label{fig:pred3}
}
\subfigure[VHM45]{%
\includegraphics[scale=0.39]{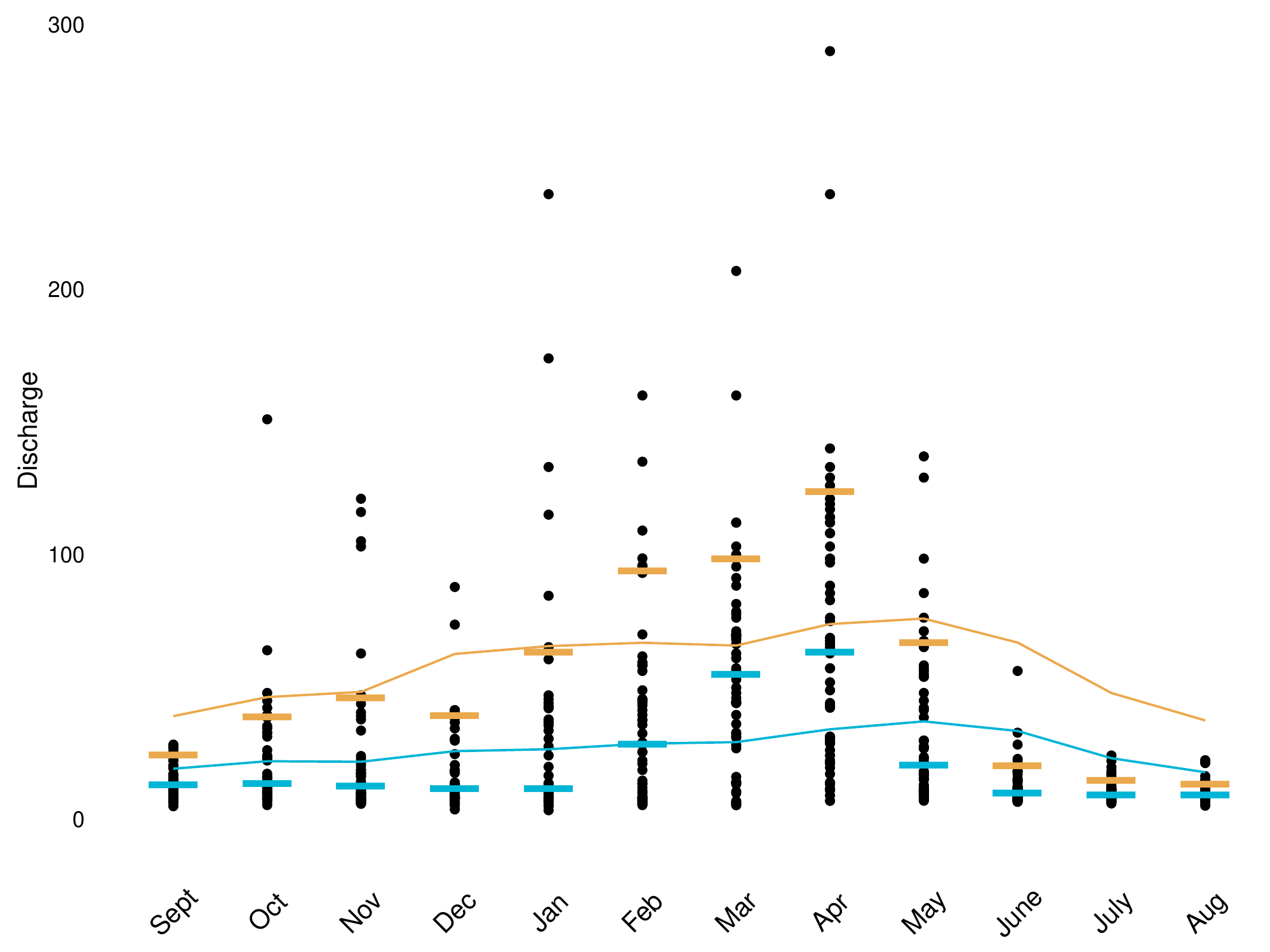}
\label{fig:pred4}
}

\caption{Predictive performance for rivers VHM10, VHM19, VHM26 and
  VHM45. 
The blue curves show predicted medians,
  while the orange curves show 90th percentile predictions. The blue bars
  show the data medians, while the orange bars show the 90th percentile
  of the data for each river. The black dots show the invidual data points. } 
\label{fig:leaveout1}
\end{figure}

\begin{figure}[!h]
\centering

\subfigure[VHM51]{%
\includegraphics[scale=0.39]{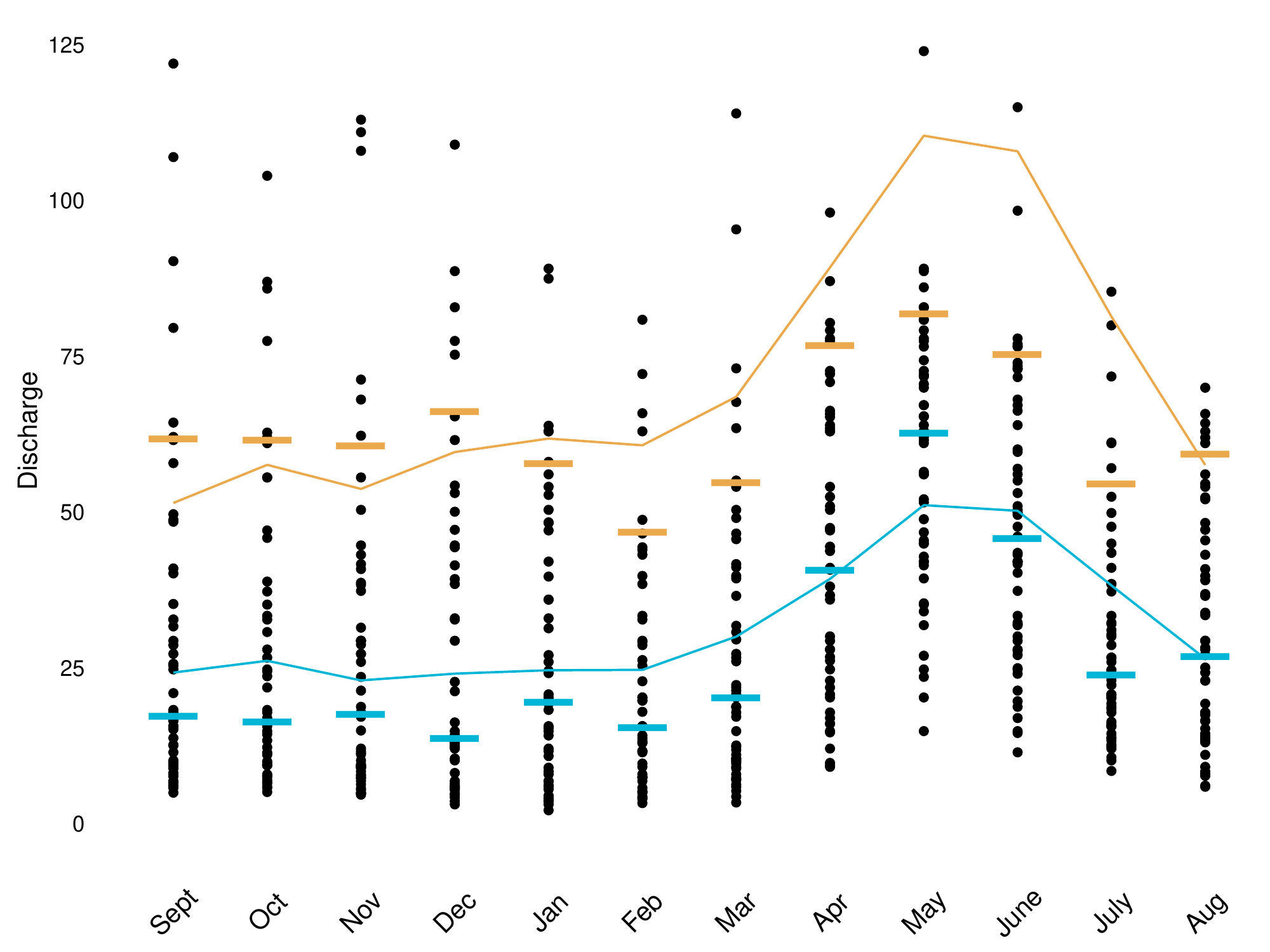}
\label{fig:pred5}
}
\subfigure[VHM198]{%
\includegraphics[scale=0.39]{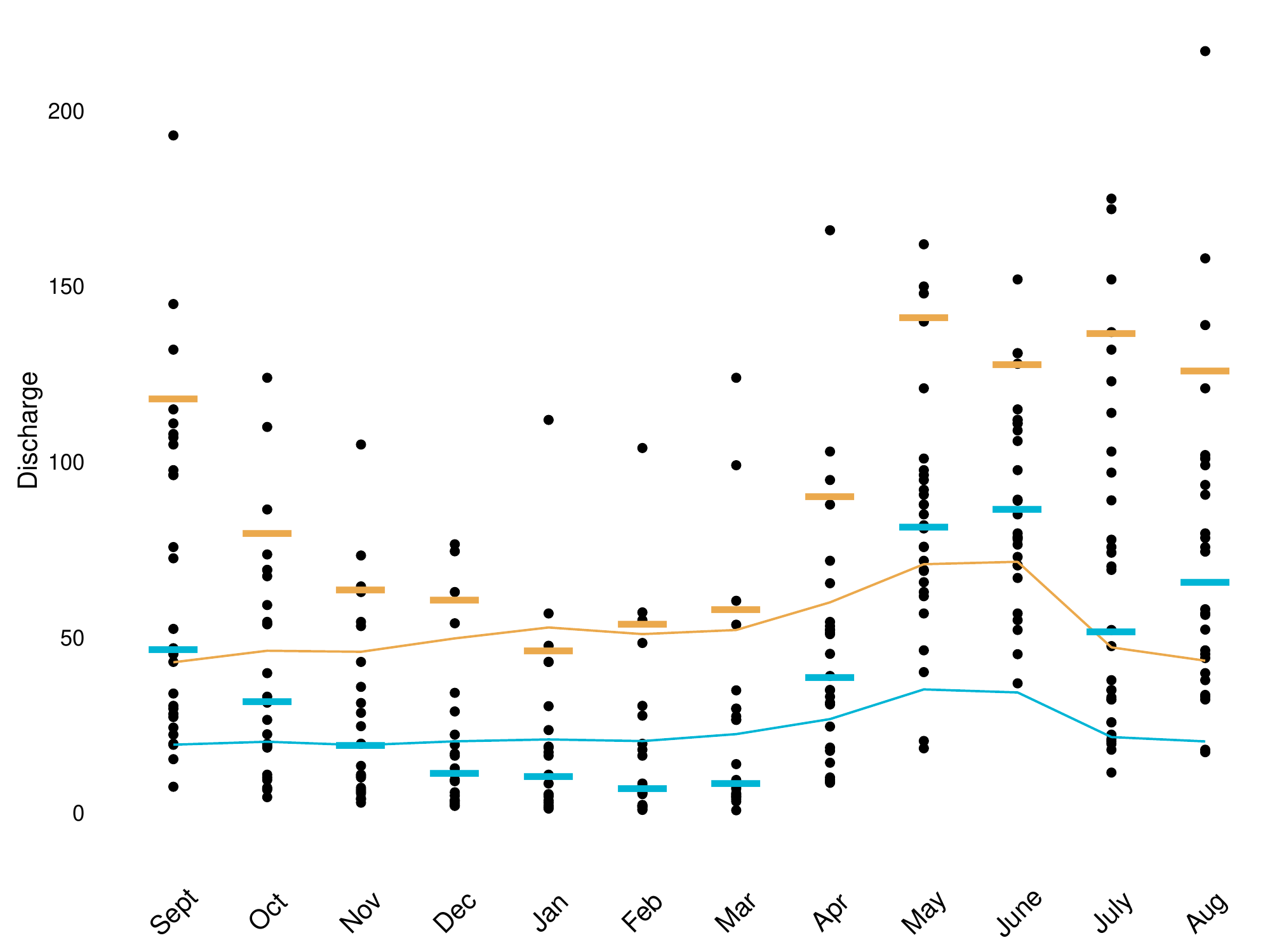}
\label{fig:pred6}
}

\subfigure[VHM200]{%
\includegraphics[scale=0.39]{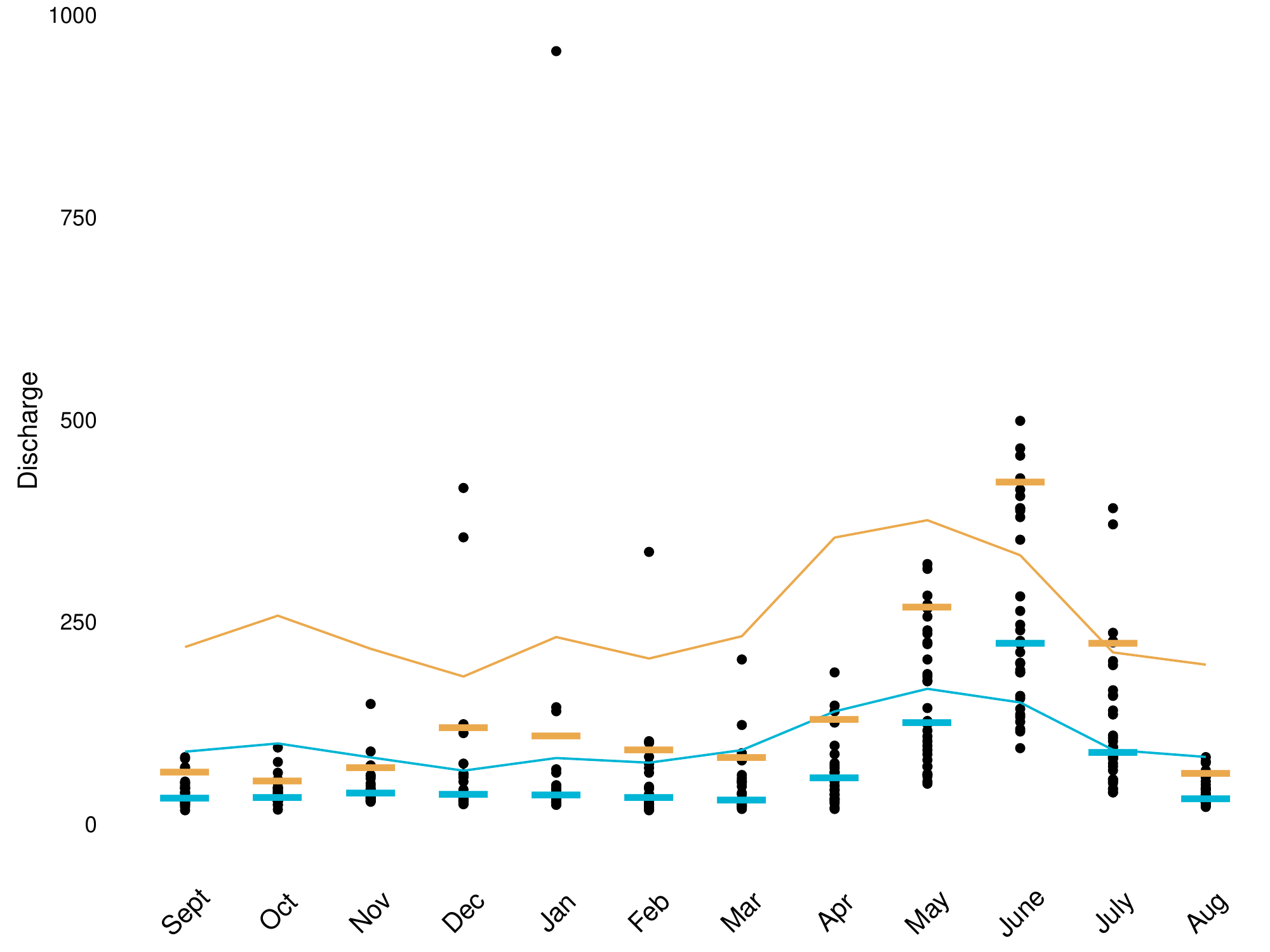}
\label{fig:pred7}
}
\subfigure[VHM204Q]{%
\includegraphics[scale=0.39]{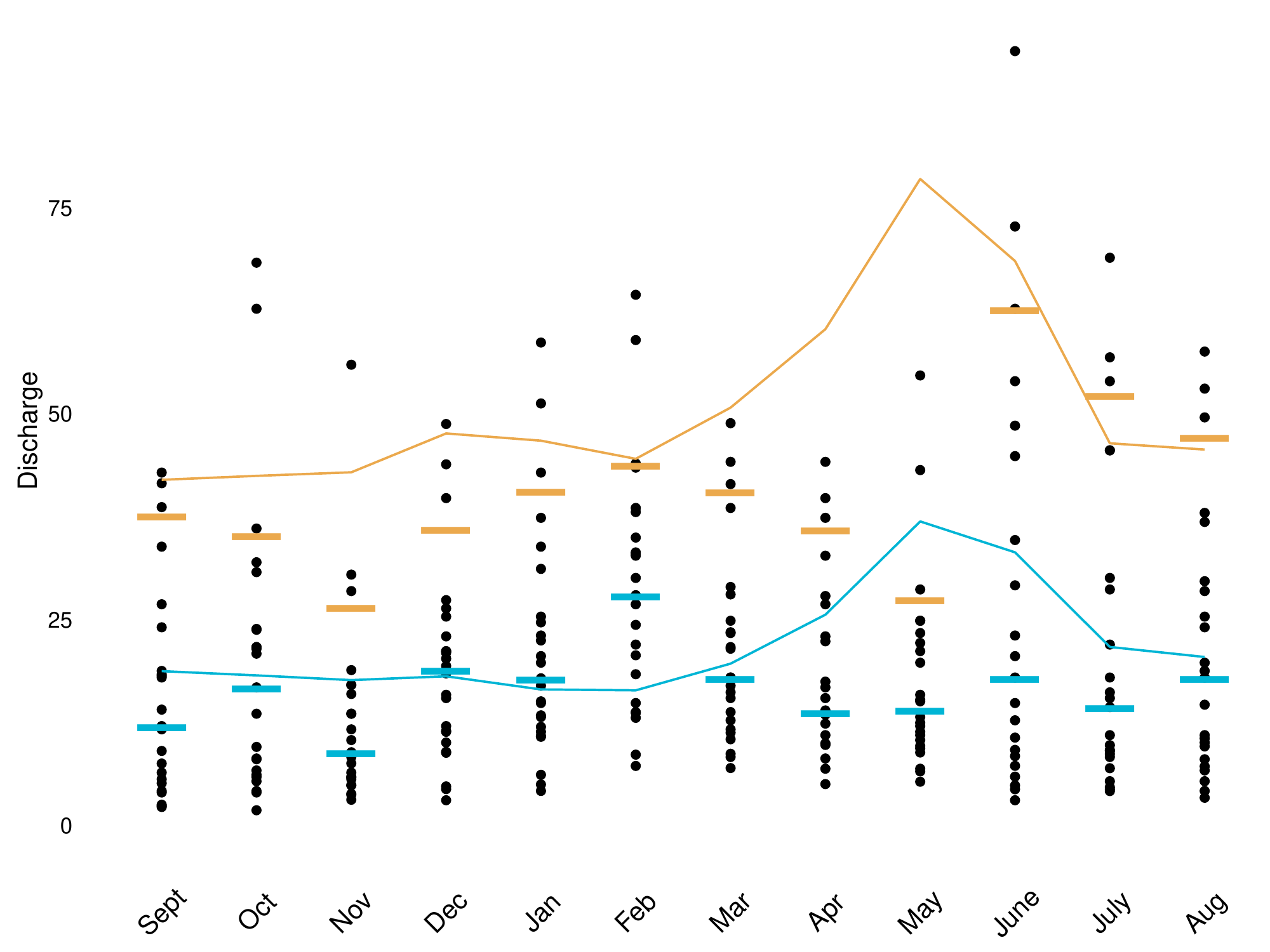}
\label{fig:pred8}
}

\caption{Predictive performance for rivers VHM51, VHM198, VHM200 and
  VHM204. 
The blue curves show predicted medians,
  while the orange curves show 90th percentile predictions. The blue bars
  show the data medians, while the orange bars show the 90th percentile
  of the data for each river. The black dots show the invidual data points. } 
\label{fig:leaveout2}
\end{figure}

Finally, we performed a cross-validation study, by leaving each river
out in turn, estimating the full model based on the remaining seven
rivers, and predicting for the left-out
river. Figures~\ref{fig:leaveout1} and \ref{fig:leaveout2} show the
results for all eight rivers. Since the aim is to predict extremes, we
do not consider prediction of the lower quantiles, but focus on the
median and the 90th percentile. The limited number of data points
(around 50) for each river-month combination 
would make estimation of higher sample quantiles such as 0.95 or 0.99 too noisy.

The model seems to predict reasonably well overall,
particularly when taking into account that the model was fitted based on
only seven river catchments, and that these are a purely out-of-sample
predictions based on sparse data. The worst prediction is for river
VHM19, which is the smaller river catchment in our data set, and is
also somewhat untypical, with smallest discharge levels overall. It is
therefore perhaps not surprising that prediction fails somewhat
here. For all the other rivers, however, the predicitive accuracy is
in our view about as good as can be expected.

%\newpage
\section{Conclusions}
\label{sec:conclusion}

We have proposed a Bayesian hierarchical model for monthly maxima of instantaneous flow. Since the number of sites is often small (as in the data used here), the ability to borrow strength between months is very important. Rather than performing twelve  (one for each month) independent linear regressions at the latent level, we fitted a linear mixed model using information jointly from all months and all sites. The use of penalised complexity priors was helpful, giving a good balance between prior information and sparse data. A thorough account of the prior elicitation for both regression coefficients and hyperparameters was given. We argue that the use of PC priors make hyperprior elicitation easier: the principle of user-defined scaling gives a useful framework for thinking about priors for hyperparameters in complex models.

Based on a preliminary analysis, it was shown that the Gumbel distribution fits the data well in most cases. However, the generalised extreme value distribution is often selected as a model for block extrema, due to its theoretic basis and it containing the Gumbel distribution as a special case. Future research on models for monthly maxima of instantaneous flow should involve assuming the generalised extreme value distribution at the data level. Assuming the same shape parameter across months would be a sensible starting point. If that is not sufficient, then assuming that each month has its own shape parameter would be a sensible extension. 

A crucial aspect of the proposed model is its capacity to predict monthly maxima of instantaneous flow at ungauged sites, provided that catchment covariates are available. The model could also be used to predict annual maxima of instantaneous flow at ungauged sites. The Bayesian approach allows for taking parameter uncertainty into account, while also helping to reduce uncertainty by using the regularising priors that are selected here. The result is reasonably good predictions compared to observed data. 
%\newpage
\section*{Acknowledgements}

We thank H{\aa}vard Rue, Andrea Riebler, Daniel Simpson and Philippe
Crochet for many helpful comments and suggestions. The data was
provided by the Icelandic Meteorological Office. The study was partly
funded by the University of Iceland Research Fund.   

%\input{Acknowledgement}

%\newpage

%\bibliographystyle{abbrv}
%\bibliographystyle{apalike}
%\bibliographystyle{authordate1}{}

\end{document}